\documentclass{emulateapj}

\begin{document}


\title{A Census of Oxygen in Star-Forming Galaxies: An Empirical Model Linking Metallicities, Star Formation Rates and Outflows}
\author{H.J. Zahid$^1$, G.I. Dima$^1$, L.J. Kewley$^{1,2}$, D.K. Erb$^3$ \& R. Dav\'{e}$^{4}$}
\affil{$^{1}$University of Hawaii at Manoa, Institute for Astronomy}
\affil{$^{2}$Australian National University, Research School of Astronomy and Astrophysics}
\affil{$^{3}$University of Wisconsin Milwaukee, Department of Physics}
\affil{$^{4}$University of Arizona, Department of Astronomy}

\begin{abstract}
In this contribution we present the first census of oxygen in star-forming galaxies in the local universe. We examine three samples of galaxies with metallicities and star formation rates at $z = 0.07, 0.8$ and $2.26$, including the SDSS and DEEP2 surveys. We infer the total mass of oxygen produced and mass of oxygen found in the gas-phase from our local SDSS sample. The star formation history is determined by requiring that galaxies evolve along the relation between stellar mass and star formation rate observed in our three samples. We show that the observed relation between stellar mass and star formation rate for our three samples is consistent with other samples in the literature. The mass-metallicity relation is well established for our three samples and from this we empirically determine the chemical evolution of star-forming galaxies. Thus, we are able to simultaneously constrain the star formation rates and metallicities of galaxies over cosmic time allowing us to estimate the mass of oxygen locked up in stars. Combining this work with independent measurements reported in the literature we conclude that the loss of oxygen from the interstellar medium of local star-forming galaxies is likely to be a ubiquitous process with the oxygen mass loss scaling (almost) linearly with stellar mass. We estimate the total baryonic mass loss and argue that only a small fraction of the baryons inferred from cosmological observations accrete onto galaxies. 

\end{abstract}


\section{Introduction}

A complete theory of galaxy formation and evolution will have to be able to self-consistently account for, among other physical processes, the star formation and chemical evolution of galaxies. Our understanding of galaxy evolution is rooted in the currently accepted cosmological model in which large-scale structure in the universe traces out the cosmic web of dark matter and growth of the universe is accelerated by dark energy. In this theoretical framework, a hierarchical formation of galaxies is favored in which larger galaxies form as the dark matter halos within which they are embedded merge over time. It remains uncertain what epoch in cosmic history this is the dominant mode of growth. However, recent observations of strong correlations observed between fundamental galaxy parameters (e.g. mass, age, size, luminosity, baryonic content and angular momentum) have lead some to question the stochastic nature of the hierarchical formation scenario \citep{Disney2008, Nair2010}. One possible resolution is that galaxies and groups of galaxies gather matter early on followed by quiescent, isolated evolution \citep{Peebles2010}. The evolution of galaxies may be simpler than a hierarchical formation model suggests.

A large number of studies have recently revealed that there exists a tight relation between stellar mass and star formation rates (SFRs) out to $z \sim 2$ \citep[among others]{Noeske2007a, Salim2007, Daddi2007, Elbaz2007, Pannella2009, Elbaz2011}. We refer to this as the MS relation. All these studies find the slope of the relation to be near unity and a $1\sigma$ scatter of $\lesssim0.3$ dex. The relation and its small scatter is taken as evidence that secular processes, such as gas accretion, are the dominant mechanism for star formation with mergers playing a minor role. In particular, \citet{Noeske2007a} suggest that the presence of an MS relation with constant scatter at several epochs implies that star formation is gradually declining with galaxies spending 67\% (95\%) of their star formation lifetime within a factor of $\sim$2 (4) of their average SFR.

Several studies have applied the observational constraints imposed by the MS relation and its evolution to uncover star formation histories of galaxies. \citet{Noeske2007b} show that their model of ``staged" galaxy evolution accounts for the observed relation. In their model, less massive galaxies have later onset of initial star formation with longer timescales of exponential decay. Similar models result if star-forming galaxies are assumed to lie on the MS relation at all epochs. Several studies have focused on this simpler approach of continuity of star formation along MS relation. \citet{Conroy2009b} combine this approach with abundance matching to dark matter halos, concluding that mergers play a minor role in mass growth of galaxies. Using this approach, \citet{Peng2010} are able to explain the shape and evolution of the observed stellar mass function for star-forming galaxies. \citet{Papovich2011} apply this technique to understand the gas accretion process at high redshifts. \citet{Leitner2011} use this technique to show that gas recycling is sufficient to fuel the observed star formation in the local universe and \citet{Leitner2012} argue that most star-forming galaxies in the local universe formed at $1<z<2$.

The chemical evolution of the gas-phase of star-forming galaxies is largely constrained by observations of the mass-metallicity (MZ) relation. \citet{Lequeux1979} were the first to show that the metallicities of galaxies increase with stellar mass. The MZ relation is well established in the local universe \citep{Tremonti2004} and has been observed for intermediate \citep{Savaglio2005, Cowie2008, Zahid2011a, Moustakas2011} and high redshift galaxies \citep{Erb2006b, Mannucci2009}. The shape of the MZ relation is observed to be relatively constant with evolution in the zero point such that galaxies at earlier redshifts are found to have lower gas-phase abundance.

The metal content of galaxies is governed by the processes of star formation and large scale gas flows. Outflowing gas has directly been observed in starburst galaxies \citep{Rupke2005, Martin2006, Tremonti2007, Rich2010, Tripp2011} and is found to be a ubiquitous phenomena in higher redshift star-forming galaxies \citep{Shapley2003, Weiner2009, Steidel2010}. A recent survey of the halos of galaxies conducted by \citet{Tumlinson2011} reveals that large reservoirs of oxygen are found to exist in the circum-galactic medium (CGM) of all star-forming galaxies. They conclude that the CGM of star-forming galaxies contains a substantial amount of gas and metals, perhaps far exceeding the gas within the galaxies themselves. In this study we use our census of oxygen to quantify the loss of metals and gas from the ISM of normal local star-forming galaxies.

Census techniques have proven to be crucial in our understanding of cosmological evolution. A well constrained inventory of the energy content of the universe is one of the greatest triumphs of modern cosmology. By comparing a census of the observed baryons in the local universe to the expected cosmological density, \citet[among others]{Fukugita1998} showed that the vast majority of baryons are not observed. This is one of the missing baryons problems. A second related problem is that the amount of baryons within galaxies is not in accord with expectations inferred from the properties of the dark matter halos within which they are embedded \citep[e.g][]{Bell2003b}. Theoretical cosmological models suggest that the missing baryons are to be found in the warm-hot intergalactic medium (WHIM) \citep{Cen1999, Dave2001}. Baryons in this phase may or may not be associated with galaxies and it remains unclear what fraction of the baryons accreted onto galaxies and were later ejected.

In this study we present a self-consistent, empirically constrained census model of oxygen in star-forming galaxies. The data used in this study are presented in Section \ref{sec:data} and the methods used in deriving the stellar masses, metallicities and SFRs of galaxies are discussed in Section \ref{sec:methods}. We parameterize the SFRs of galaxies as a function of stellar mass and redshift in Section \ref{sec:msr} by examining the MS relation at several redshifts. We describe the various components of our oxygen census in Section \ref{sec:inventory}. We develop our self-consistent empirical models for the star formation and chemical history of galaxies in Section \ref{sec:mdot} and \ref{sec:mzr}, respectively, by imposing the continuity condition that galaxies build up their stellar mass by evolving along the empirical relation between stellar mass and SFR with the metallicity inferred from the MZ relation at several redshifts. In Section \ref{sec:census} we present the results of our census and in Section \ref{sec:uncertainties} we discuss systematics and uncertainties in our approach. We provide a discussion of our results in Section \ref{sec:discussion} and a summary of our results in Section \ref{sec:summary}. For this study we adopt the standard cosmology $(H_{0}, \Omega_{m}, \Omega_{\Lambda}) = (70\, km\, s^{-1}\, Mpc^{-1}, 0.3, 0.7)$.

\section{The Data and Sample Selection}
\label{sec:data}

In this section we describe the local sample of galaxies from SDSS (Section 3.1), our intermediate redshift sample from DEEP2 (Section 3.2) and a high redshift sample from \citet[Section 3.3]{Erb2006b}. Galaxies are primarily selected such that the chemical properties can be determined from their spectra. The binned data are given in a Table \ref{tab:data}.

\subsection{The SDSS Sample}

We draw our local sample from the SDSS DR7 which consists of $\sim$900,000 galaxies spanning a redshift range of $0 < z < 0.7$ \citep{Abazajian2009}. The survey has a Petrosian limiting magnitde of $r_P$ = 17.8 covering 8,200 $deg^2$. The spectra have a nominal spectral range of 3900 - 9100$\mathrm{\AA}$ and a spectral resolution of R $\sim$ 2000. We make use of the $ugriz$-band photometry  available for each object \citep{Stoughton2002} and the publicly released emission line fluxes measured by the MPA-JHU group\footnote{http://www.mpa-garching.mpg.de/SDSS/DR7/}. We refer to the sample presented here as the ``metallicity selected SDSS sample''.

We correct for dust extinction in the emission lines by inferring a reddening correction from the Balmer decrement. For case B recombination with electron temperature T$_e$  = 10$^4$K and electron density $n_e = 10^2$ cm$^{-3}$, the intrinsic H$\alpha$/H$\beta$ ratio is expected to be 2.86 \citep{Osterbrock1989}. We get the intrinsic color excess, E(B$-$V), and the correction for dust attenuation using the extinction law of \citet{Cardelli1989} and a corresponding $R_v = 3.1$. We note that the results of this study are not dependent on our choice of a particular extinction law.

We select a pure star-forming sample of local emission line galaxies from the SDSS DR7. We first distinguish star-forming galaxies from AGN by constraining the ionizing radiation source using the [OIII]$\lambda5007$, [NII]$\lambda6584$, H$\beta$ and H$\alpha$ emission lines \citep{Baldwin1981, Veilleux1987, Kauffmann2003, Kewley2006}. In particular, we remove galaxies using the equation given in \citet{Kewley2006} where 
\begin{equation}
\mathrm{log([OIII]/H\beta)} > 0.61/\left(\mathrm{log([NII]/H\alpha)} - 0.05 \right) + 1.3.
\end{equation}
In order to avoid aperture effects, we require a g-band fiber aperture covering fraction $>30\%$ in addition to imposing a lower redshift limit of 0.04 \citep{Kewley2004}. The median covering fraction for the metallicity selected SDSS sample is $38\%$. \citet{Kewley2006} find that the SDSS sample is incomplete at higher redshifts and in order to minimize evolutionary effects we also impose an upper limit redshift cutoff of $z = 0.1$. 

In order to establish comparable samples, galaxies in the local sample are selected from SDSS using the same selection criteria as the DEEP2 sample. In particular, galaxies are selected to have S/N of H$\beta > 3$, $\sigma_{R23} < 2$ and equivalent width of H$\beta > 4\mathrm{\AA}$. Here, $\sigma_{R23}$ is the error in the $R23$ parameter which is the ratio of the oxygen nebular emission ([OII]$\lambda3727$ doublet and [OIII]$\lambda4959, 5007$) to H$\beta$. These particular selection criteria gives us a sample of $\sim$20,000 star-forming galaxies in the limited redshift range of $0.04<z<0.1$.

\subsection{The DEEP2 Sample}

Our sample of intermediate redshift star-forming galaxies is taken from the DEEP2 survey \citep{Davis2003}. Details of sample selection and properties are given in \citet{Zahid2011a}, here we summarize the data selection. The survey consists of $\sim$45,000 galaxies targeted mostly in the redshift range of $0.7 > z > 1.4$ by applying a color preselection using $BRI$-band photometry \citep{Coil2004}. The survey has a limiting magnitude of $R_{AB} = 24.1$ and covers 3.5 $deg^2$. The spectra have a nominal spectral range of 6500 - 9100$\mathrm{\AA}$ and a resolution of R $\sim$ 5000. The emission line equivalent widths are measured in \citet{Zahid2011a} and we adopt the same values here. 

The sample selection is based on the spectral and photometric properties of the galaxies. We reduce AGN contamination by first requiring that log$(R23) < 1$ (Kewley et al., in prep). Using the color separation for blue and red galaxies parameterized by \citet{Willmer2006}, \citet{Weiner2007} conclude that only a small fraction of blue galaxies in DEEP2 appear to harbor AGN whereas a large fraction of red emission line galaxies show evidence of AGN emission. We further limit AGN contamination by removing 48 galaxies in the sample that are classified as red galaxies using the color division of \citet{Willmer2006}. 

Given the nominal spectral coverage, the redshift range of galaxies in our sample is limited by the necessity to simultaneously observe both the [OII]$\lambda3727$ doublet and the [OIII]$\lambda5007$ emission lines which are required for chemical analysis. We further require that the S/N H$\beta > 3$, the error of the $R23$ emission line ratio, $\sigma_{R23}$, be less than 2 and that equivalent width of H$\beta > 4\mathrm{\AA}$. Finally, due to ambiguity in the metallicity determination of DEEP2 galaxies at low stellar masses (see Section 3.2) we also remove all galaxies with $M_\ast < 10^{9.2}M_\odot$. This selection criteria gives us a sample of 1348 star-forming galaxies in the limited redshift range of $0.75<z<0.82$.

\subsection{The E06 Sample}

\citet[E06 hereafter]{Erb2006b} determine the MZ relation from 87 star-forming galaxies at $z \sim 2.2$ selected from a larger sample of 114 galaxies described in \citet{Erb2006a}. The galaxies are selected on the basis of their rest-frame UV colors and redshifts are determined from their UV spectra. All galaxies in the sample have $UGRK$-band photometry. Most also have $J$-band photometry and 32 galaxies have been observed at 3.6, 4.5, 5.4 and 8.0 $\mu$m with IRAC onboard the Spitzer Space Telescope. H$\alpha$ spectra were obtained using NIRSPEC on the Keck II telescope.

The metallicity for these galaxies is determined from the emission line ratio of [NII]$\lambda6584$ to H$\alpha$. The [NII]$\lambda6584$ line is sensitive to metallicity with the strength of the line decreasing with decreasing metallicity. The S/N of the individual galaxy spectra is insufficient to measure the weak [NII]$\lambda6584$ line. In order to increase the S/N of their spectra and to increase the chance of detecting [NII] emission line at low metallicites, E06 stack 14 or 15 individual galaxy spectra binned by stellar mass into 6 composite spectra. The [NII]$\lambda6584$ and H$\alpha$ emission line fluxes and metallicities are measured from these composite spectra.

\begin{deluxetable}{cccc}
\tablecaption{Data}
\tablehead{\colhead{log$(M_\ast/M_\odot)$} & \colhead{12 + log(O/H)} &\colhead{E(B$-$V)} & \colhead{log($\Psi$)} }
\startdata
&~~~~~~~~~~~~~~~~~~~SDSS&& \\
\hline
8.51 & 8.707 $\pm$ 0.004 & 0.11 $\, \pm \, $ 0.02     &       $\!\!$-0.20 $\, \pm \, $ 0.02\\
8.82 & 8.736 $\pm$ 0.006 & 0.13 $\, \pm \, $ 0.02     &       $\!\!$-0.12 $\, \pm \, $ 0.02\\
8.97 & 8.787 $\pm$ 0.007 & 0.15 $\, \pm \, $ 0.02     &       $\!\!$-0.19 $\, \pm \, $ 0.02\\
9.08 & 8.819 $\pm$ 0.008 & 0.16 $\, \pm \, $ 0.02     &       $\!\!$-0.20 $\, \pm \, $ 0.02\\
9.17 & 8.859 $\pm$ 0.007 & 0.18 $\, \pm \, $ 0.02     &       $\!\!$-0.19 $\, \pm \, $ 0.02\\
9.23 & 8.875 $\pm$ 0.006 & 0.19 $\, \pm \, $ 0.01     &       $\!\!$-0.14 $\, \pm \, $ 0.01\\
9.30 & 8.900 $\pm$ 0.006 & 0.20 $\, \pm \, $ 0.01     &       $\!\!$-0.12 $\, \pm \, $ 0.01\\
9.36 & 8.920 $\pm$ 0.006 & 0.22 $\, \pm \, $ 0.01     &       $\!\!$-0.10 $\, \pm \, $ 0.01\\
9.41 & 8.923 $\pm$ 0.006 & 0.23 $\, \pm \, $ 0.01     &       $\!\!$-0.07 $\, \pm \, $ 0.01\\
9.45 & 8.946 $\pm$ 0.006 & 0.24 $\, \pm \, $ 0.01     &       $\!\!$-0.06 $\, \pm \, $ 0.01\\
9.49 & 8.947 $\pm$ 0.006 & 0.24 $\, \pm \, $ 0.01     &       $\!\!$-0.03 $\, \pm \, $ 0.01\\
9.54 & 8.969 $\pm$ 0.006 & 0.27 $\, \pm \, $ 0.01     &       $\!\!$-0.02 $\, \pm \, $ 0.01\\
9.57 & 8.977 $\pm$ 0.004 & 0.26 $\, \pm \, $ 0.01     &       $\!\!$-0.01 $\, \pm \, $ 0.01\\
9.61 & 8.993 $\pm$ 0.006 & 0.27 $\, \pm \, $ 0.01     &       0.03 $\, \pm \, $ 0.01\\
9.64 & 8.989 $\pm$ 0.005 & 0.28 $\, \pm \, $ 0.01     &       0.05 $\, \pm \, $ 0.01\\
9.68 & 9.007 $\pm$ 0.004 & 0.28 $\, \pm \, $ 0.01     &       0.07 $\, \pm \, $ 0.01\\
9.71 & 9.010 $\pm$ 0.004 & 0.30 $\, \pm \, $ 0.01     &       0.09 $\, \pm \, $ 0.01\\
9.75 & 9.022 $\pm$ 0.004 & 0.30 $\, \pm \, $ 0.01     &       0.11 $\, \pm \, $ 0.01\\
9.78 & 9.035 $\pm$ 0.003 & 0.32 $\, \pm \, $ 0.01     &       0.13 $\, \pm \, $ 0.01\\
9.81 & 9.037 $\pm$ 0.004 & 0.33 $\, \pm \, $ 0.01     &       0.16 $\, \pm \, $ 0.01\\
9.85 & 9.048 $\pm$ 0.003 & 0.33 $\, \pm \, $ 0.01     &       0.17 $\, \pm \, $ 0.01\\
9.88 & 9.056 $\pm$ 0.003 & 0.36 $\, \pm \, $ 0.01     &       0.22 $\, \pm \, $ 0.01\\
9.92 & 9.059 $\pm$ 0.003 & 0.36 $\, \pm \, $ 0.01     &       0.24 $\, \pm \, $ 0.01\\
9.95 & 9.068 $\pm$ 0.003 & 0.38 $\, \pm \, $ 0.01     &       0.27 $\, \pm \, $ 0.01\\
10.00 & 9.061 $\pm$ 0.003 & 0.39 $\, \pm \, $ 0.01     &       0.32 $\, \pm \, $ 0.01\\
10.04 & 9.081 $\pm$ 0.002 & 0.40 $\, \pm \, $ 0.01     &       0.34 $\, \pm \, $ 0.01\\
10.09 & 9.084 $\pm$ 0.003 & 0.43 $\, \pm \, $ 0.01     &       0.40 $\, \pm \, $ 0.01\\
10.15 & 9.088 $\pm$ 0.002 & 0.44 $\, \pm \, $ 0.01     &       0.47 $\, \pm \, $ 0.01\\
10.24 & 9.086 $\pm$ 0.003 & 0.47 $\, \pm \, $ 0.01     &       0.53 $\, \pm \, $ 0.01\\
10.39 & 9.095 $\pm$ 0.002 & 0.52 $\, \pm \, $ 0.01     &       0.73 $\, \pm \, $ 0.01\\
\hline
&~~~~~~~~~~~~~~~~~~~DEEP2&& \\
\hline
9.25     &       8.69 $\, \pm \, $ 0.02     &       0.17 $\, \pm \, $ 0.03     &       0.35 $\, \pm \, $ 0.03\\
9.32     &       8.76 $\, \pm \, $ 0.02     &       0.19 $\, \pm \, $ 0.02     &       0.32 $\, \pm \, $ 0.02\\
9.39     &       8.78 $\, \pm \, $ 0.02     &       0.20 $\, \pm \, $ 0.03     &       0.37 $\, \pm \, $ 0.03\\
9.44     &       8.77 $\, \pm \, $ 0.02     &       0.21 $\, \pm \, $ 0.03     &       0.39 $\, \pm \, $ 0.03\\
9.49     &       8.74 $\, \pm \, $ 0.01     &       0.21 $\, \pm \, $ 0.03     &       0.43 $\, \pm \, $ 0.03\\
9.56     &       8.80 $\, \pm \, $ 0.02     &       0.23 $\, \pm \, $ 0.03     &       0.49 $\, \pm \, $ 0.03\\
9.64     &       8.83 $\, \pm \, $ 0.02     &       0.24 $\, \pm \, $ 0.04     &       0.54 $\, \pm \, $ 0.04\\
9.72     &       8.84 $\, \pm \, $ 0.02     &       0.26 $\, \pm \, $ 0.03     &       0.60 $\, \pm \, $ 0.03\\
9.79     &       8.86 $\, \pm \, $ 0.02     &       0.27 $\, \pm \, $ 0.05     &       0.60 $\, \pm \, $ 0.05\\
9.87     &       8.92 $\, \pm \, $ 0.02     &       0.31 $\, \pm \, $ 0.05     &       0.72 $\, \pm \, $ 0.05\\
9.97     &       8.94 $\, \pm \, $ 0.02     &       0.33 $\, \pm \, $ 0.03     &       0.77 $\, \pm \, $ 0.03\\
10.07     &       8.93 $\, \pm \, $ 0.02     &       0.35 $\, \pm \, $ 0.03     &       0.82 $\, \pm \, $ 0.03\\
10.18     &       8.96 $\, \pm \, $ 0.01     &       0.38 $\, \pm \, $ 0.03     &       0.92 $\, \pm \, $ 0.03\\
10.33     &       9.00 $\, \pm \, $ 0.01     &       0.43 $\, \pm \, $ 0.02     &       0.98 $\, \pm \, $ 0.02\\
10.59     &       9.04 $\, \pm \, $ 0.01     &       0.53 $\, \pm \, $ 0.04     &       1.29 $\, \pm \, $ 0.04\\

\hline
&~~~~~~~~~~~~~~~~~~E06&& \\
\hline
9.14     &       $<$8.55     &       0.14 $\, \pm \, $ 0.09     &       1.40 $\, \pm \, $ 0.14\\

9.56     &       8.72 $\, \pm \, $ 0.07     &       0.21 $\, \pm \, $ 0.07     &       1.41 $\, \pm \, $ 0.05\\
9.89     &       8.82 $\, \pm \, $ 0.06     &       0.28 $\, \pm \, $ 0.08     &       1.60 $\, \pm \, $ 0.06\\
10.12     &       8.86 $\, \pm \, $ 0.06     &       0.33 $\, \pm \, $ 0.07     &       1.52 $\, \pm \, $ 0.08\\
10.32     &       8.92 $\, \pm \, $ 0.06     &       0.39 $\, \pm \, $ 0.09     &       1.78 $\, \pm \, $ 0.06\\
10.73     &       8.97 $\, \pm \, $ 0.05     &       0.52 $\, \pm \, $ 0.06     &       1.96 $\, \pm \, $ 0.06\\
\enddata
\label{tab:data}
\tablecomments{The stellar mass (column 1), metallicity determined from the diagnostic of \citet[column 2]{Kobulnicky2004}, fitted E(B$-$V) from Equation \ref{eq:ebvfit} (column 3) and SFR(column 4) for the SDSS, DEEP2 and E06 samples. For the E06 sample the metallicity has been converted from \citet{Pettini2004} to \citet{Kobulnicky2004} using the conversion constants in \citet{Kewley2008}. DEEP2 and E06 sample we have corrected for dust extinction when determining SFRs from the Balmer lines using the E(B$-$V) values given in column 3. For the DEEP2 and SDSS data the values are the median in bins of stellar mass. The errors are determined from bootstrapping and are analogous to the standard error on the mean. For the E06 sample, the errors are the standard error of the mean. Each data bin is equally populated such that the SDSS, DEEP2 and E06 samples contain $\sim700$, $\sim90$ and $\sim15$ galaxies in each bin, respectively. Electronic version available upon request.}
\end{deluxetable}

\section{Methods}
\label{sec:methods}
In this section we discuss our methods for determining stellar masses (Section 4.1), metallicities (Section 4.2) and SFRs (Section 4.3).

\subsection{Stellar Mass}
We use the Le Phare\footnote{$\url{http://www.cfht.hawaii.edu/{}_{\textrm{\symbol{126}}}arnouts/LEPHARE/cfht\_}$ $\url{lephare/lephare.html}$} code developed by Arnouts, S. \& Ilbert, O. to estimate the galactic stellar mass. This code estimates the stellar masses of galaxies by comparing photometry with stellar population synthesis models in order to determine the mass-to-light ratio which is then used to scale the observed luminosity \citep{Bell2003b, Fontana2004}. We synthesize magnitudes from the stellar templates of \citet{Bruzual2003} and use a \citet{Chabrier2003} IMF. The 27 models have two metallicities and seven exponentially decreasing star formation models (SFR $\propto e^{-t/\tau}$) with $\tau = 0.1,0.3,1,2,3,5,10,15$ and $30$ Gyrs. We apply the extinction law of \citet{Calzetti2000} allowing E(B$-$V) to range from 0 to 0.6 and the stellar population ages range from 0 to 13 Gyrs. The median statistical error for the derived stellar masses, determined from propagating the uncertainty in the photometry, is 0.15 dex. Though systematic effects may be large \citep{Drory2004, Conroy2009a}, we have consistently measured the stellar masses for our different samples giving us a robust relative measure \citep[R. Swindle, private comm.]{Swindle2011}.

We account for emission line contributions by taking the \citet{Kennicutt1998b} relation between the synthesized UV luminosity and SFR and the emission lines. This treatment accounts for H$\alpha$, H$\beta$ and [OII]$\lambda3727$ and [OIII]$\lambda4959, 5007$ \citep{Ilbert2009b}. In our sample, making no correction for the emission line contributions does not significantly alter our mass determinations. We adopt the median of the mass distribution and take the 68\% confidence interval as a measure of the error. In \citet{Zahid2011a} we compare this method with the method used by the MPA/JHU group to determine stellar masses of the SDSS galaxies. We find that the two estimates differ by a constant offset of $\sim$0.2 dex and that the dispersion between the two methods is 0.14 dex.

\begin{figure}
\begin{center}
\includegraphics[width=\columnwidth]{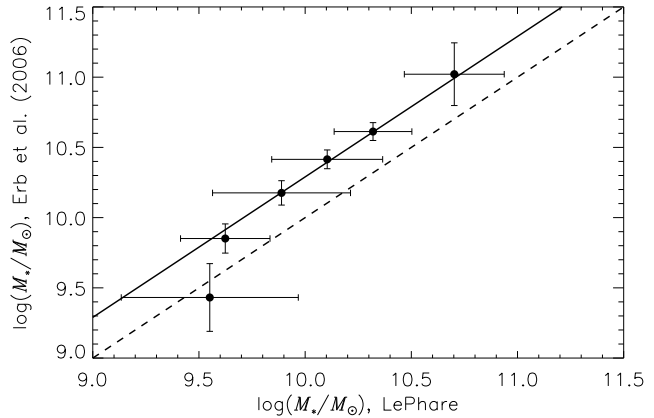}
\end{center}
\caption{The stellar mass determined by \citet{Erb2006c} plotted against our determination using Le Phare. The dashed line is the one-to-one agreement and the solid line is offset by 0.29 dex. In the five higher mass bins, the stellar mass estimates used in E06 are greater by a factor of 2 (0.29 dex).}
\label{fig:mass_erb}
\end{figure}

E06 measure stellar masses using a similar method of comparing photometry with stellar population synthesis models \citep{Erb2006c}. However, an important difference is that they measure the ``total" stellar mass, which is the integral of the SFR over the lifetime of the galaxy. They find that this stellar mass is $\sim10 - 40\%$ higher than the instantaneous (or what they term ``current living") stellar mass. However, most mass estimates found in the literature are generally the instantaneous and not ``total" stellar mass. We recalculate the stellar masses for their individual galaxies and bin the data according to their original binning. In Figure \ref{fig:mass_erb} we compare the mass estimates for each of their 6 bins. The x-axis shows our mass estimate and y-axis is the original mass estimate from E06. The error bar represents the rms dispersion of stellar masses in each mass bin. For the five higher mass bins, we find that there is a constant offset such that the E06 estimates are 0.29 dex higher than our estimate. For the lowest mass bin, there appears to be a near one-to-one agreement, though the 0.29 dex offset is within the errors. This is most likely due to the fact that the lowest mass galaxies are younger and therefore have a much closer agreement between their instantaneous and ``total" stellar mass. Moreover, we do not concern ourselves too much with the lowest mass bin as the metallicity measure is only an upper limit for this bin. Since we do not rebin the E06 data in determining the metallicity we adopt their stellar mass values but subtract 0.29 dex to make them consistent with our estimates.

\begin{figure*}
\begin{center}
\includegraphics[width=1.8\columnwidth]{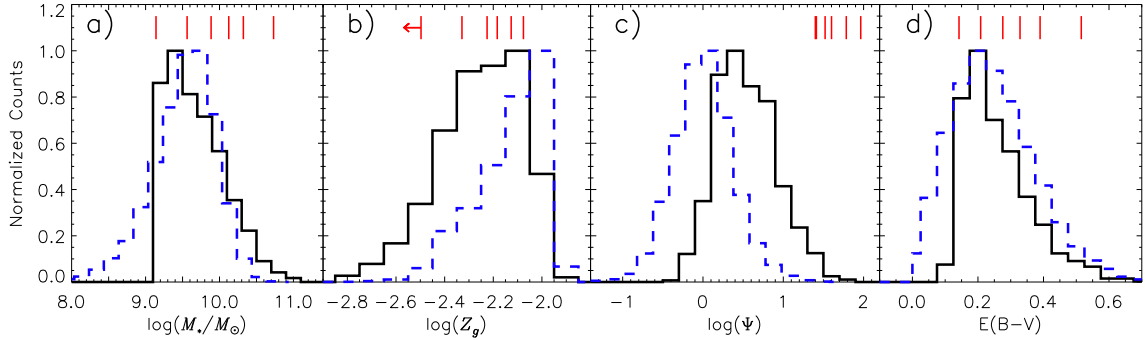}
\end{center}
\caption{A histogram of the a) stellar mass, b) metallicity, c) SFR ($M_\odot$ yr$^{-1}$) and d) the fitted E(B$-$V) for the DEEP2 (solid black) and SDSS (dashed blue) samples. The values for the 6 binned data points of E06 are shown by the red ticks.}
\label{fig:hist}
\end{figure*}

Figure \ref{fig:hist}a shows the distribution of the stellar masses for all three samples.

\subsection{Metallicity}

We use two strong-line methods to determine the metallicity of galaxies in our sample. In this section we only present the parameterization of the calibration and defer a detailed discussion of the methods until Section 8.3. For the SDSS and DEEP2 sample we determine metallicities using the calibration of \citet{Kobulnicky2004}. This method relies on the $R23$ and $O32$ parameters which are defined as
\begin{equation}
R23 = \frac{\mathrm{[OII]}\lambda3727 + \mathrm{[OIII]}\lambda4959, 5007}{\mathrm{H}\beta}
\end{equation}
and 
\begin{equation}
O32 = \frac{\mathrm{[OIII]}\lambda4959, 5007}{\mathrm{[OII]}\lambda3727}.
\end{equation}
Here the ratio implies the ratio of the measured line intensities. We have used the assumption that the ratio of the fluxes of [OIII]$\lambda5007$ to [OIII]$\lambda4959$ is 3 \citep{Osterbrock1989}. Due to the higher S/N of the [OIII]$\lambda5007$ line, for the sum of the [OIII]$\lambda4959$ and [OIII]$\lambda5007$ flux we adopt a value of 1.33 times the [OIII]$\lambda5007$ flux.

The $R23$ strong-line method is known to be sensitive to the ionization parameter. The ionization parameter characterizes the ionization state of the gas and quantitatively represents the number of ionizing photons per second per unit area divided by the hydrogen density. The ionization parameter can be constrained by measuring the ratio of the line intensity of the same element at two ionization stages. The method of \citet{Kobulnicky2004} does this explicitly by using the $O32$ line ratio. Because both the metallicity and ionization parameter are interdependent, an iterative scheme is used, the details of which are provided in the appendix of \citet{Kewley2008}. In particular we use equations A4 and A6.

The DEEP2 data are not flux calibrated so we use the line equivalent widths with a correction applied for Balmer absorption \citep{Zahid2011a}. For consistency we determine the SDSS metallicities using line equivalent widths as well. Several studies have established that line equivalent widths can be substituted for line fluxes when measuring metallicities if data are not flux calibrated or a reliable reddening estimate is unavailable \citep{Kobulnicky2003b, Moustakas2010, Zahid2011a}. In particular, we test this on our SDSS sample and find that the dispersion between the metallicities measured using equivalent widths and dereddened line fluxes is $\sim$0.05 dex which is less than intrinsic uncertainties of the strong-line method. The average difference in the MZ relation derived using the two methods is $<0.01$ dex.

One issue with using the $R23$ method is that metallicity is not a monotonic function of $R23$. For a particular value of the $R23$ parameter there are two possible values of metallicity known as the upper and lower branch metallicity. This degeneracy is generally alleviated by employing a second metallicity diagnostic that relies on line ratios that are monotonic with metallicity. For our SDSS sample of galaxies, the use of [NII]/$H\alpha$ reveals that the vast majority ($\sim$99\%) of galaxies are on the upper branch. For our DEEP2 sample, the nominal wavelength coverage of the spectra does not allow us to simultaneously observe the $R23$ lines and [NII]/H$\alpha$. We assume all galaxies with $M_\ast>10^{9.2}M_\odot$ lie on the upper metallicity branch \citep[for more details see Figure 5 in][]{Zahid2011a}. 

The calibration of \citet{Kobulnicky2004} is based on theoretical photoionization models. Empirical methods rely on calibrating strong-line ratios to metallicities determined using temperature sensitive auroral lines. E06 determine metallicities using the empirical calibration of \citet{Pettini2004}. The metallicity is given by 
\begin{equation}
\mathrm{12+log(O/H)} = 8.90 + 0.57 \times N2
\label{eq:met_pettini}
\end{equation}
where $N2$ = log([NII]$\lambda6584$/H$\alpha$). Using $\sim28,000$ galaxies from the SDSS DR4, \citet{Kewley2008} derive constants for converting between various diagnostics. We use these conversion constants to consistently determine metallicities for all three samples using both the theoretical and empirical calibrations of \citet{Kobulnicky2004} and \citet{Pettini2004}.

The metallicity of a galaxy is traditionally defined as a number density of oxygen relative to hydrogen and is given as 12 + log(O/H). For this study we require the mass density of oxygen relative to hydrogen. We convert number density to mass density using the relation
\begin{eqnarray}
\mathrm{log(}Z_g) &= &12 + \mathrm{log(O/H)} - \\ 
&& 12 - \mathrm{log} \left( \frac{M_O/M_H}{XM_H + YM_{He}} \right) \nonumber \\
& = & \mathrm{log(O/H)} - \mathrm{log} \left( \frac{15.999/1.0079}{0.75\cdot1.0079 + 0.25\cdot4.0026} \right). \nonumber
\label{eq:mass_density}
\end{eqnarray}
For the remainder of the paper, $Z_g$ refers to the gas-phase \emph{mass} density of oxygen relative hydrogen. Also, when referring to metallicity or gas-phase oxygen abundance we mean to refer to $Z_g$.

Figure \ref{fig:hist}b shows the distribution of the metallicities for all three samples.

\subsection{Star Formation Rate}


The SFRs for the SDSS sample are derived by the MPA/JHU group using the technique of \citet{Brinchmann2004} with additional improvements given by \citet{Salim2007}. The strong emission lines of each galaxy are fit using the nebular emission models of \citet{Charlot2001}. All emission lines contribute to constraining the dust attenuation, though the largest contribution comes from the H$\alpha$/H$\beta$ ratio. At the median redshift of the SDSS, about a 1/3 of the galaxy light is contained within the 3" fiber. In order to account for losses, \citet{Brinchmann2004} derive an aperture correction which was improved upon by \citet{Salim2007}. We convert from a Kroupa to Chabrier IMF by dividing by 1.06.

\begin{figure}
\begin{center}
\includegraphics[width=\columnwidth]{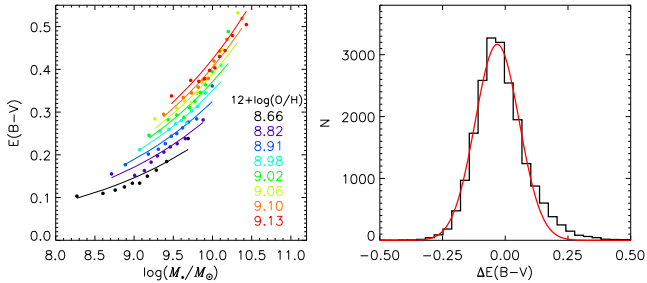}
\end{center}
\caption{A fit to E(B$-$V) as function of stellar mass and metallicity. The left panel shows the E(B$-$V) sorted into 8 color-coded bins of metallicity and plotted as function of stellar mass. The underlying curves are the fitted relation given by Equation \ref{eq:ebvfit} and color-coded by metallicity. The right panel shows a histogram of the residuals between the observed E(B$-$V) and those determined from the fitted relation.}
\label{fig:ebv_fit}
\end{figure}

We estimate the SFRs for the DEEP2 and E06 sample from the H$\beta$ and H$\alpha$ emission lines, respectively. In order to reliably estimate the SFR, it is necessary to make a correction to the Balmer line flux due to dust extinction. Since the nominal wavelength coverage of both the DEEP2 and E06 spectra do not allow us to observe multiple order Balmer lines, we are not able to derive a correction from the Balmer decrement. \citet{Garn2010b} have shown that for a sample of star-forming galaxies from the SDSS the extinction determined from the H$\alpha$/H$\beta$ ratio can be predicted from the stellar mass, H$\alpha$ luminosity or metallicity of the galaxy with a dispersion of $\sim$0.1 dex. The relation between stellar mass and dust extinction found by \citet{Garn2010b} is shown to be valid out to $z \sim 1.5$ \citep{Sobral2011} and a relation between dust extinction, stellar mass and metallicity is also observed in galaxies at $z\sim2$ \citep{Reddy2010}.

\citet{Xiao2012} obtain a slightly better fit (dispersion of 0.07 dex) by incorporating galaxy inclination and using a different parameterization. We determine E(B$-$V) for our sample of SDSS galaxies from the Balmer decrement assuming the \citet{Cardelli1989} extinction law and parameterize E(B$-$V) as a function of stellar mass and metallicity using a similar formulation to \citet{Xiao2012}. The color excess is given by
\begin{equation}
\mathrm{E(B-V)} = (p_{0} + p_{1}  Z^{p_2}) \times M^{p_3},
\label{eq:ebvfit}
\end{equation}
where $Z = 10^{(12 + \mathrm{log(O/H)} - 8)}$ and $M = M_{\ast}/10^{10}$. The metallicities are derived using the calibration of \citet{Kobulnicky2004}. We perform a non-linear least square fit using the MPFIT\footnote{http://purl.com/net/mpfit} set of routines \citep{Markwardt2009}. The data are weighted by their 1$\sigma$ statistical uncertainties and the errors are propagated to the parameters. The fit returns $p_{0} = 0.12 \pm 0.01$,  $p_{1} = 0.041 \pm 0.006$, ${p_2} = 0.77 \pm 0.06$ and ${p_3} = 0.240 \pm 0.002$. The left panel of Figure \ref{fig:ebv_fit} shows the mean E(B$-$V) binned by stellar mass and color coded by metallicity. The fit to these binned data are shown by the color coded curves. The right panel shows the residuals between the observed and fitted color excess. There is a slight tail to the distribution owing to the lower quality of the fit to the high stellar mass, high metallicity and high color excess data. The rms dispersion between the observations and the fit is $\sim$0.11 dex.

We measure the SFRs for the DEEP2 sample of galaxies using the H$\beta$ luminosity. In order to obtain a flux calibration, we integrate the DEEP2 spectra over the $I$-band filter response and compare to the observed $I$-band magnitude. This gives us an estimate for the conversion between counts on the detector and flux given in ergs s$^{-1}$. This procedure accounts for both throughput and slit losses assuming that the line-to-continuum emission ratio is constant inside and outside of the slit. This assumption is reasonable as slit losses are generally small because the vast majority of DEEP2 galaxies have effective radius smaller than the 1" slit width \citep{Weiner2007}. We linearly fit the flux calibrated continuum in a 80$\mathrm{\AA}$ window about the H$\beta$ emission line and scale the equivalent width by the continuum to obtain a H$\beta$ flux. 

We use the extinction law of \citet{Cardelli1989} and the fitted E(B$-$V) values given by Equation \ref{eq:ebvfit} to correct the observed H$\beta$ flux in the DEEP2 sample for dust attenuation. We obtain the luminosity from the flux using the relation $L(\mathrm{H}\beta) = F_{\mathrm{H}\beta} \times 4 \pi D_{L}^2$ where $F_{\mathrm{H}\beta}$ is the dereddened H$\beta$ flux and $D_{L}$ is the luminosity distance determined from the redshift. We determine the SFR from the \citet{Kennicutt1998b} relation between SFR and H$\alpha$ luminosity given by 
\begin{equation}
\mathrm{SFR} = 7.9 \times 10^{-42} L(\mathrm{H}\alpha) [\mathrm{ergs} \,\, \mathrm{s}^{-1}].
\label{eq:ha_sfr}
\end{equation}
We take $L(\mathrm{H}\alpha) = 2.86 \times L(\mathrm{H}\beta)$ and scale the SFR from Salpeter to Chabrier IMF by dividing by 1.7.

E06 infer their SFRs from the H$\alpha$ luminosity corrected for dust extinction using the E(B$-$V) determined from SED fitting. A factor of 2 aperture correction was applied and the SFRs were determined using the \citet{Kennicutt1998b} relation converted to a Chabrier IMF. We redetermine the SFRs for their 87 galaxies by applying an extinction correction determined from Equation \ref{eq:ebvfit} to their observed H$\alpha$ luminosities. Following E06, we apply a factor of 2 aperture correction to the observed luminosity. We determine the SFR from Equation \ref{eq:ha_sfr} and bin data according to E06.

Figure \ref{fig:hist}c shows the distribution of the SFRs for all three samples.

\section{Galactic Stellar Mass Growth}
\label{sec:msr}
\subsection{The Observed Relation between Stellar Mass and SFR}

\begin{figure}
\begin{center}
\includegraphics[width=\columnwidth]{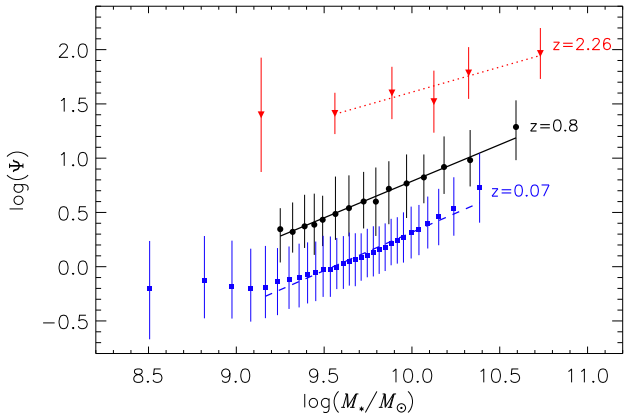}
\end{center}
\caption{The logarithm of the SFR ($M_\odot$ yr$^{-1}$) plotted in bins of stellar mass for the SDSS (blue squares), DEEP2 (black circles) and E06 (red triangles) samples. For the SDSS and DEEP2 sample the median SFR is plotted and the error bars indicate the interval containing 68\% of the data. For the E06 sample the median is plotted and the error bars are the standard deviation of the distribution. The dashed blue, solid black and dotted red lines are fits to the to the observed MS relation for the SDSS, DEEP2 and E06 samples, respectively. The fit parameters are given in Table \ref{tab:sfr}.}
\label{fig:sfr}
\end{figure}

In Figure \ref{fig:sfr} we plot the MS relation for the SDSS (blue squares), DEEP2 (black circles) and E06 (red triangles) star-forming galaxies at redshifts of $z = 0.07, 0.78$ and 2.26, respectively. For the SDSS, DEEP2 and E06 samples the median SFR is determined in bins of stellar masses. For the SDSS, DEEP2 and E06 samples there are $\sim700$, $\sim90$ and $\sim15$ galaxies in each data bin. The flattening in the relation observed at lower stellar masses is due to incompleteness (see below) such that the lower SFR galaxies fall out of our selected samples at lower stellar masses. Down to log($M_\ast) \sim 9.2$ for the SDSS and DEEP2 sample and log($M_\ast) \sim 9.5$ for the E06 sample the relation between stellar mass and SFR is well described by a power law. 

For the SDSS and DEEP2 samples the error bars plotted on the data points indicate the interval containing 68\% of the data within each mass bin. For the E06 sample the error bars indicate the standard deviation of the distribution within each mass bin. The median scatter in the SFRs is 0.25, 0.27 and 0.24 dex for the SDSS, DEEP2 and E06 samples, respectively. The scatter in the three samples is comparable and is roughly constant with stellar mass.  The observed scatter in our samples is comparable to the $\lesssim 0.3$ dex found in studies examining the $M_\ast$-SFR relation out to $z\sim2$ \citep{Salim2007, Elbaz2007, Noeske2007a, Daddi2007, Pannella2009}.

\begin{deluxetable}{lccc}
\tablewidth{210pt}
\tablecaption{Mass-SFR Relation}
\tablehead{\colhead{Sample} & \colhead{redshift} &\colhead{$\delta$} & \colhead{$\gamma$} } 

\startdata
SDSS & 0.07 & 0.317 $\pm$ 0.003 & 0.71 $\pm$ 0.01 \\
DEEP2 & 0.78 & 0.787 $\pm$ 0.009 & 0.67 $\pm$ 0.02  \\
E06 & 2.26 & 1.608 $\pm$ 0.028 & 0.46 $\pm$ 0.07  \\
\enddata
\label{tab:sfr}
\tablecomments{The fitted MS relation for our three samples as parameterized by Equation \ref{eq:sfr_slope}.}
\end{deluxetable}

We fit a linear relation between log($\Psi$) and log($M_\ast$/$M_\odot$) in the mass range where the samples are monotonically declining using the \emph{linfit.pro} routine in IDL. The form of the fit is given by
\begin{equation}
\mathrm{log(SFR)} = \delta + \gamma \cdot [ \mathrm{log}(M_\ast/M_\odot) - 10],
\label{eq:sfr_slope}
\end{equation}
where $\delta$ is the logarithm of the SFR at $10^{10} M_\odot$ and $\gamma$ is the power law index. The dashed blue, solid black and dotted red lines in Figure \ref{fig:sfr} are the fits to the SDSS, DEEP2 and E06 samples, respectively. We give the fit parameters and the errors in Table \ref{tab:sfr}. The errors for the SDSS and DEEP2 fit parameters are bootstrapped. The errors for the E06 sample are determined from propagating the standard error of each data bin, $\sigma/\sqrt{n}$, where $\sigma$ is the standard deviation of the distribution and $n$ is the number of objects in each bin. The power law index of the relation is consistent (within the small errors) for the SDSS and DEEP2 samples. We have applied a factor of 2 aperture corrections to all galaxies in the E06 sample. This correction is likely overcorrecting the lower mass galaxies which could explain the shallower slope observed in the MS relation. The zero point of the MS relation shows evolution to higher values with increasing redshift.

\subsection{Sample Incompleteness}
\begin{figure}
\begin{center}
\includegraphics[width=\columnwidth]{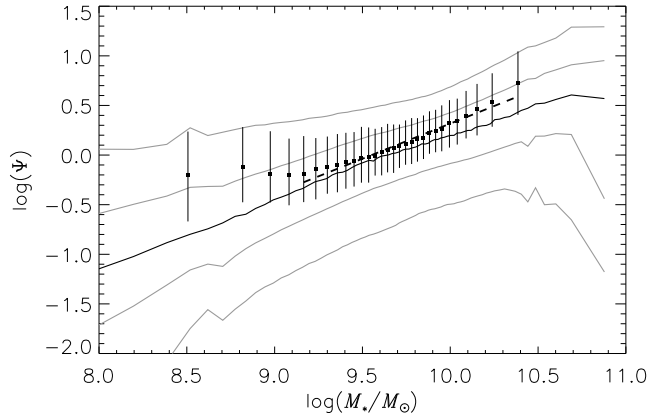}
\end{center}
\caption{The SFR ($M_\odot$ yr$^{-1}$) plotted against stellar mass for our metallicity selected SDSS sample (black points) along with larger comparison sample from SDSS. The black points and error bars are the observed MS relation and its scatter for our metallicity selected SDSS sample. The dashed black line is our fitted relation given by Equation \ref{eq:sfr_slope} with parameters given in Table \ref{tab:sfr}. The solid black curve is the MS relation for our larger comparison sample and the gray lines are the 68 and 95\% contours.}
\label{fig:sfr_inc}
\end{figure}

In all three of our samples the MS relation flattens out at lower stellar masses (see Figure \ref{fig:sfr}). The galaxies in all three of the samples are selected such that chemical properties can be determined from the spectra. Determining chemical properties from the emission lines of star-forming galaxies generally requires high S/N observations and therefore samples with determined metallicities are more susceptible to incompleteness. We demonstrate that the observed flattening is due to incompleteness by comparing our sample of $\sim$20,000 galaxies for which we have well determined metallicities with a larger, more complete sample of SDSS star-forming galaxies. 

We select the larger comparison sample by first removing AGN using the [OIII]/H$\beta$ vs. [NII]/H$\alpha$ diagram \citep{Kauffmann2003, Kewley2006} and imposing the similar redshift restrictions ($z < 0.1$) as our metallicity selected SDSS sample. This gives us a parent sample $\sim$200,000 star-forming galaxies. The aperture corrected SFRs given in the SDSS DR7 rely on a Bayesian technique principally fitting H$\alpha$ and H$\beta$ lines to determine extinction corrected SFRs \citep{Brinchmann2004, Salim2007}. We select galaxies from the parent sample by requiring a S/N of 5 in the H$\alpha$ and H$\beta$ emission lines. This gives us a sample of $\sim$140,000 star-forming galaxies with $z < 0.1$ and detections of H$\alpha$ and H$\beta$. We note that the distribution of SFRs in our comparison sample are largely independent of choice of S/N cuts and therefore the observed incompleteness in our metallicity selected SDSS sample likely results from our limit on the equivalent width of H$\beta$. 

Figure \ref{fig:sfr_inc} shows the SFRs for both our metallicity selected SDSS sample of galaxies and the larger comparison sample. The black data points with error bars and dashed line are the same as in Figure \ref{fig:sfr}. The solid black line is the median SFRs of our comparison sample sorted into 75 bins of stellar mass and the gray curves are the 68 and 95\% contours of the data. The MS relation in the larger comparison sample has a slope of 0.65 which is sightly more shallow than our metallicity selected SDSS sample. This slope is consistent with the slope determined by \citet{Salim2007} for MS relation of SDSS galaxies. The zero point of the larger sample is slightly lower as compared to our metallicity selected SDSS sample. Over the range where we fit the MS relation, the difference in the SFRs for the two samples is $< 0.1$ dex and both relations show similar scatter. 

From this comparison it is clear that the flattening observed in the MS relation in Figure \ref{fig:sfr} is due to incompleteness. The slightly higher zero point of our metallicity selected SDSS sample suggests that there is a slight bias against low star-forming galaxies over all mass ranges. This is consistent with our expectation that the higher S/N and equivalent width of H$\beta$ required in the spectra for chemical analysis selects against galaxies with weak lines and low levels of star formation. Despite this selection effect, Figure \ref{fig:sfr_inc} demonstrates that incompleteness does not significantly bias our determination of the MS relation.

\citet{Noeske2007a} examine the MS relation for DEEP2 galaxies in the redshift range of $0.2 < z < 1$. They determine SFRs from the 24 $\mu$m flux and add in the SFRs determined from the DEEP2 emission lines with no extinction correction to account for unobscured star formation. They determine the fit to the MS relation in the range where their sample is $>\!95\%$ complete. The MS relation determined by \citet[but see \citet{Dutton2010}]{Noeske2007a} at $z = 0.78$ is consistent with our determination for our DEEP2 sample at the same redshift, despite the fact that our sample is selected differently and we have determined SFRs from extinction corrected emission lines. Our correction for dust extinction relies on the correlation between extinction, stellar mass and metallicity (Section 4.3). This necessarily introduces a correlation between stellar mass and SFR. The fact that the MS determined from our DEEP2 sample is consistent within the errors to the relation fit by \citet{Noeske2007a} implies that our dust correction is not introducing a \emph{spurious} correlation between stellar mass and SFR. Furthermore, we conclude that the MS relation we determine at $z = 0.78$ from our DEEP2 sample is not biased due to incompleteness.

The completeness of the E06 sample at $z = 2.26$ is much more uncertain. This sample is UV-selected which may bias against star-forming galaxies with low levels of star formation if they are present at this redshift. However, we note that the observed relation determined from the E06 sample is consistent within the errors to the zero point and scatter found by \citet{Daddi2007} using a highly complete sample at $z\sim2$ and \citet{Pannella2009} who use radio stacking to determine SFRs. This suggests that the observed MS relation in the E06 sample is not significantly biased.

\subsection{Parameterization of SFR}

\begin{figure}
\begin{center}
\includegraphics[width=\columnwidth]{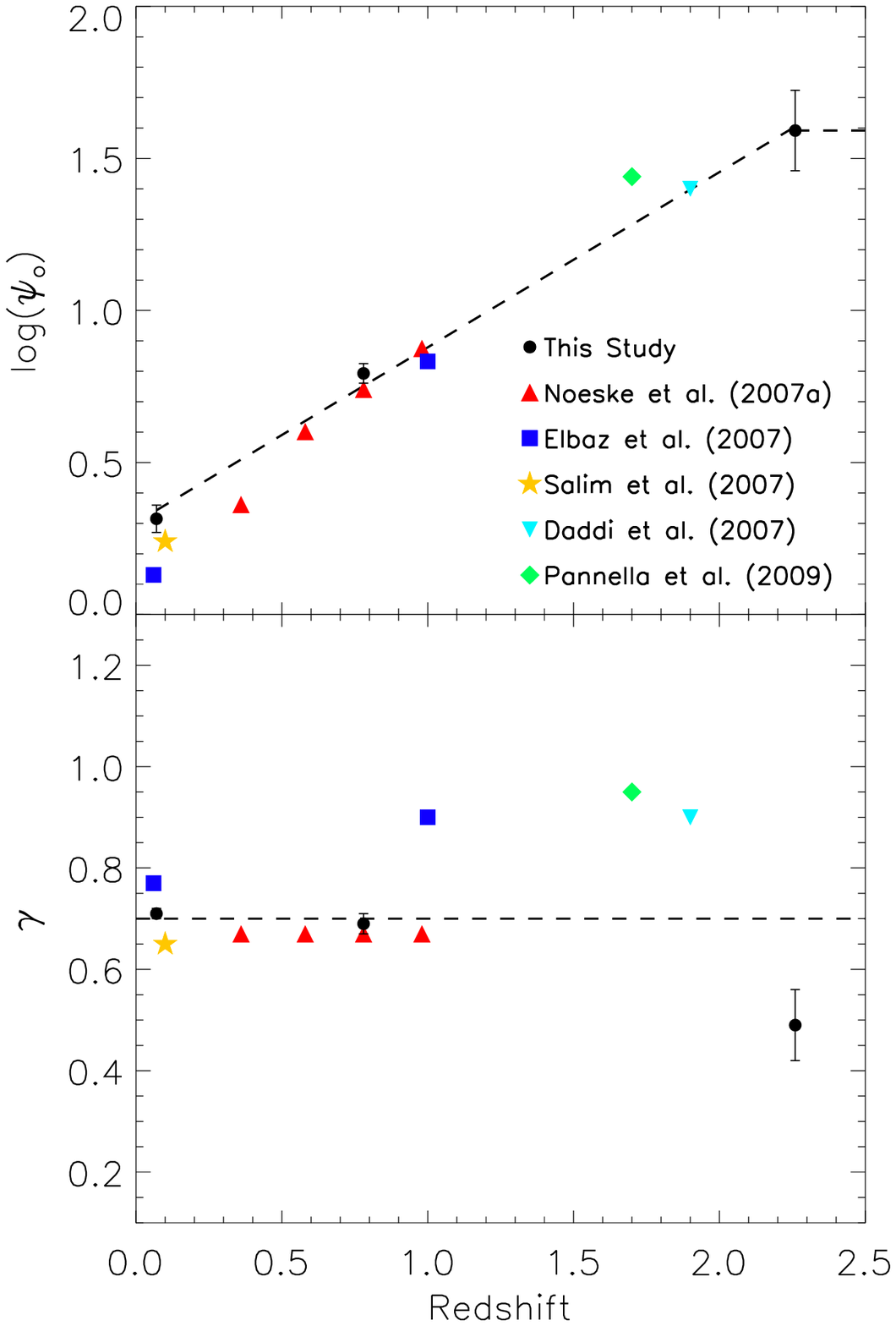} 
\end{center}
\caption{The SFR ($M_\odot$ yr$^{-1}$) for a $10^{10} M_\odot$ galaxy ($\psi_o$, top panel) and the power law index of the MS relation ($\gamma$, bottom panel) as a function of redshift for our data (black points) alongside several values from the literature. The data come from \citet[red triangles]{Noeske2007a}, \citet[blue squares]{Elbaz2007}, \citet[yellow star]{Salim2007}, \citet[cyan triangle]{Daddi2007} and \citet[green diamond]{Pannella2009}. The dashed line in the top panel is the fit for $\psi_o$ as a function of redshift for our three samples given by Equation \ref{eq:psi0}. The dashed line in the bottom panel is our adopted value of 0.7 for the power law index of the MS relation, $\gamma$.}
\label{fig:ft}
\end{figure}

With a well defined MS relation at several redshifts, we can parameterize the SFR as a function of mass and redshift. We seek a parameterization where the variables of mass and redshift are separated such that the SFR is given by
\begin{equation}
\Psi(M_\ast, z) = \psi_{o}(z) \cdot f(M_\ast).
\label{eq:sfr}
\end{equation}
Here $\Psi(M_\ast, z)$ is the SFR as a function of stellar mass and redshift, $\psi_{o}(z)$ is the zero point which evolves with redshift and $f(M_\ast)$ is the relation between stellar mass and SFR. This definition explicitly assumes that the slope of the relation between stellar mass and SFR has no time dependency.

As discussed in Section 4.1, the SFR as a function of stellar mass is well fit by a power law in stellar mass with an index near unity. We therefore parameterize the stellar mass dependency as 
\begin{equation}
f(M_\ast) = (M_\ast/10^{10} M_\odot)^{\gamma}, 
\label{eq:fm}
\end{equation}
where the only free parameter, $\gamma$, is the power law index of the relation. With this definition for $f(M_\ast)$, the zero point, $\psi_o(z)$, is the SFR of a $10^{10} M_\odot$ galaxy. The zero point of the relation as a function of redshift, as we show below, is well fit by an exponential and we parameterize it as 
\begin{equation}
\psi_o(z) = \alpha  \cdot \mathrm{exp}(\beta z).
\label{eq:psi0}
\end{equation}
Here, $\alpha$ is the zero point at $z = 0$ and $\beta$ determines the rate of exponential increase with redshift.

Figure \ref{fig:ft} shows a compilation taken from the literature for the values of $\gamma$ (bottom panel) and $\psi_o$ (top panel) as defined in Equations \ref{eq:fm} and \ref{eq:psi0}, respectively. For comparison, data from this study are plotted as black points on these plots. We have converted the stellar masses and SFRs for the data shown in Figure \ref{fig:ft} such that they are consistent with the \citep{Chabrier2003} IMF. The data selection and method of SFR  determination vary. In the local universe, \citet{Salim2007} derive SFRs by fitting the UV/optical broad-band SED. \citet{Elbaz2007} adopt the \citet{Brinchmann2004} aperture corrected SFRs given in the DR4. At $0.2 < z < 1$, the measurements of \citet{Noeske2007a}\footnote{We have taken the parameters of the \citet{Noeske2007a} MS relation from the compilation of \citet{Dutton2010}.} have SFRs determined from a combination of uncorrected emission line fluxes and 24 $\mu$m fluxes. At $z\sim1$, \citet{Elbaz2007} determine SFRs from the infrared luminosity inferred from the 24 $\mu$m flux. At $z\sim2$, \citet{Daddi2007} determine SFRs from dust corrected UV luminosities and \citet{Pannella2009} determine SFRs from stacking analysis of the radio continuum.

Despite the differences in methodology, our determination of the zero point, $\psi_o$ (top panel Figure \ref{fig:ft}), is remarkably consistent with values found in the literature. In the local universe, we find a marginally higher zero point, which is due to incompleteness (see Section 4.2). At $z = 0.78$ we determine the same relation (within the errors) as \citet{Noeske2007a}. At $z = 2.26$ the relation determined from the E06 sample is consistent with the higher redshift data. Out to $z \sim 1$, the slope of the MS relation, $\gamma$ (bottom panel of Figure \ref{fig:ft}), are consistent in most of the studies. At $z > 1$ the slopes show some variation, though the variation is only about $\pm$35\%. We note that in this study, our results are not sensitive to the adopted value of $\gamma$ because the redshift evolution of the SFR is significantly greater than the magnitude of the variation introduced by a $\pm35\%$ change in $\gamma$.

In order to derive $\Psi(M_\ast, z)$, we need to determine values for the three free parameters, $\alpha, \beta$ and $\gamma$. For the SDSS and DEEP2 samples the slopes are consistent (within the errors) and we adopt a constant value of 0.7 for $\gamma$ in Equation \ref{eq:fm} \citep[see also][]{Whitaker2012}. This is shown by the dashed line in the bottom panel of Figure \ref{fig:ft}. To determine $\alpha$ and $\beta$ we fit a line of the form 
\begin{equation}
\mathrm{log}[\psi_o(z)] = \mathrm{log}(\alpha) + 0.434 \, \beta z,
\label{eq:sfh}
\end{equation}
to our data. The fit is shown by the dashed line in the top panel of Figure \ref{fig:ft}. We note that we have fit only to our data, but the fit is more or less consistent with all the studies compiled from the literature. At $z>2.26$ we assume that $\psi_o$ is flat noting that our results are not strongly dependent on this extrapolation because in our model most of the stellar mass growth in star-forming galaxies in the local universe occurs at $z<2.26$ \citep[see also][]{Leitner2012}.

\begin{figure}
\begin{center}
\includegraphics[width=\columnwidth]{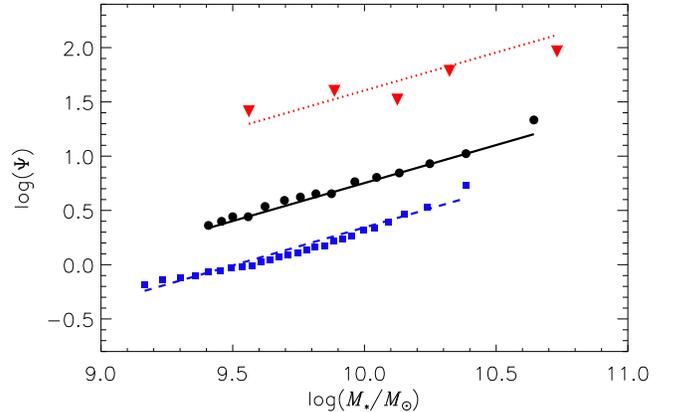}
\end{center}
\caption{The fitted relation for the SFR ($M_\odot$ yr$^{-1}$) as function of redshift, $\Psi(M_\ast, z)$, given by Equation \ref{eq:sfr_fit} plotted over our three samples.}
\label{fig:sfr_fit}
\end{figure}

With both $\psi_o(z)$ and $f(M_\ast)$ determined, we can parameterize the SFRs of star-forming galaxies as a function of stellar mass and redshift by entering the fitted results back into Equation \ref{eq:sfr}. The SFR as a function of stellar mass and redshift is given by
\begin{equation}
\Psi(M_\ast, z) = 2.00 \cdot \mathrm{exp}(1.33 z) \left( \frac{M_\ast}{10^{10}}\right)^{0.7} [M_\odot \, \mathrm{yr}^{-1}].
\label{eq:sfr_fit}
\end{equation}
In Figure \ref{fig:sfr_fit} we show the fitted relation plotted over the observations. Though the fitted slope of the MS relation at $z = 2.26$ was shallower (see Table \ref{tab:sfr}), the data is still reasonably well described by the relation given in Equation \ref{eq:sfr_fit}.

\section{Inventory of Oxygen}
\label{sec:inventory}

In this study we compare an estimate of the total amount of oxygen produced with the amount found in the ISM and stars of local star-forming galaxies. We focus on oxygen because it is a primary element produced in massive stars that end their lives as Type II supernovae. Oxygen is the most abundant element produced and therefore the easiest to measure from strong nebular emission lines originating in the ISM of galaxies. Oxygen production depends only weakly on the initial metallicity \citep{Thomas1998, Kobayashi2006} and is considered a robust tracer of total metal production. We assume that oxygen production is independent of metallicity. The rate of oxygen production is then simply given by
\begin{equation}
\frac{dM^{o}_{T}}{dt} = P_o \, \Psi.
\label{eq:mtodot}
\end{equation}
Here $M^{o}_{T}$ is the mass of oxygen produced in solar mass units, $P_o$ is a constant representing the mass of newly synthesized oxygen per solar mass of gas converted to stars and $\Psi$ is the SFR. We can estimate the total oxygen production by integrating Equation \ref{eq:mtodot}. This is given by
\begin{equation}
M^{o}_{T}(t > t_i) = P_o \, \int^t_{t_i} \Psi(t^\prime) \,dt^\prime.
\label{eq:mzint}
\end{equation}
The amount of oxygen produced between some initial time, $t_i$, and some later time, $t$, is proportional to the total amount of star formation during that epoch. This equation implies that $M_T^o \propto M_\ast$, where $M_T^o$ is the \emph{total} amount of oxygen produced in a galaxy and $M_\ast$ is the stellar mass of the galaxy. $\Psi$ is explicitly parameterized as a function of redshift (see Equation \ref{eq:sfr_fit}) which we convert into a function time by assuming the conversion between redshift and time using the IDL routine \emph{galage.pro}.

The second component of our oxygen census is the oxygen found in the ISM. We define the global gas-phase oxygen mass abundance such that $Z_g = M_g^o/M_g$. Here, $M_g$ and $M_{g}^{o}$ are the mass of gas and the mass of oxygen in the gas-phase, respectively, given in solar mass units. The total oxygen mass in the gas-phase is given by rewriting this equation such that $M_{g}^{o} = Z_g \,M_g$. Both $Z_g$ and $M_g$ are measured quantities.

The final component of our census of oxygen in star-forming galaxies is the oxygen locked up in stars. Assuming that the gas in galaxies is well mixed and mass return from stars is instantaneous, the rate at which oxygen is locked up in stars is given by
\begin{equation}
\frac{dM_{\ast}^{o}}{dt} = (1 - R) \, Z_g \, \Psi.
\label{eq:szodot}
\end{equation}
Here $M_{\ast}^{o}$ is the mass of oxygen going into stars in solar mass units and $R$ is a constant representing the fraction of gas that is returned to the ISM through various stellar mass loss processes. We can integrate Equation \ref{eq:szodot} to get the total amount of oxygen locked up in low mass stars. This is given by
\begin{equation}
M^{o}_{\ast}(t>t_i) = (1 - R)\, \int^t_{t_i} Z_g(t^\prime) \, \Psi(t^\prime) \,dt^\prime.
\end{equation}
$M^o_\ast$ represents the amount of oxygen in the ISM that gets forever locked up in low mass stars in the time between $t_i$ and $t$. Similar to $\Psi$, $Z_g$ is a explicitly a function of redshift which we convert into a function of time. In Sections \ref{sec:mdot} and \ref{sec:mzr} we develop an empirically constrained, self-consistent approach for determining $\Psi(t)$ and $Z_g(t)$, respectively.

\section{Tracing Galaxies through Cosmic Time}
\label{sec:mdot}
One interpretation of the MS relation and its evolution is that most star formation is driven largely by secular processes such as gas accretion and mergers do not play a significant role \citep[e.g.][]{Noeske2007a, Dutton2010, Elbaz2011}. Under the assumption that stellar mass build up is a purely secular process, we can trace stellar mass evolution of galaxies by assuming that 
\begin{equation}
\frac{dM_\ast}{dt} = (1 - R) \,\Psi.
\label{eq:mdot}
\end{equation}
Here $\frac{dM_\ast}{dt}$ is the time rate of change of stellar mass, $R$ is the return fraction which accounts for gas that is formed into stars but then returned to the ISM via various mass loss processes (e.g. supernovae and stellar winds) and $\Psi$ is the SFR. For simplicity we assume that R is a constant (see Section \ref{sec:uncertainties} and the Appendix for more discussion). 

If we further assume the continuity condition that star-forming galaxies \emph{evolve along the MS relation} and SFRs are a (determined) function of stellar mass and redshift (see Equation \ref{eq:sfr_fit}), it is straightforward to integrate Equation \ref{eq:mdot} in order to determine the stellar mass as a function of redshift. This is given by
\begin{equation}
M_\ast(t) = M_{\ast, i} + (1 - R) \, \int^t_{t_i} \Psi (M_\ast,\, t^\prime) \,dt^\prime.
\label{eq:mint}
\end{equation}
Here $M_\ast(t)$ is the stellar mass at some redshift, $t < t_i$. $M_{\ast, i}$ is the stellar mass at some initial time, $t_i$, and the integral of $\Psi (M_\ast, t)$ represents the gain in stellar mass, modulo the return fraction $R$, from some initial time $t_i$ to some later time $t$. If $M_{\ast,i}$ is set to an arbitrarily low value ($10^6 M_\odot$) and the upper limit of the integral is the current epoch, then $t_i$ is simply the formation time. This is similar to the staged model of \citet{Noeske2007b} such that lower stellar mass star-forming galaxies in the local universe begin forming their stars at later times.

Equation \ref{eq:mint} is a simple, yet powerful tool for identifying star-forming galaxies at lower redshifts with their progenitors at higher redshifts under the assumption of secular evolution (i.e. no contribution from merging) and that the SFRs are a known function of stellar mass and redshift (see Equation \ref{eq:sfr_fit}); the latter being observationally determined. Similar approaches have been developed in the literature and we direct the reader to \citet{Leitner2012} for a detailed discussion of the implications of such a model for stellar mass growth. Because of the stellar mass dependence of $\Psi$ and the fact that SFR is measured in units of solar masses per year and not solar masses per redshift, Equation \ref{eq:mdot} does not have a straightforward analytical solution and thus requires numerical integration. The integration of Equation \ref{eq:mdot} gives both the stellar mass and star formation history of the galaxy. 

\section{Galactic Chemical Evolution}
\label{sec:mzr}
\begin{figure*}
\begin{center}
\includegraphics[width=1.5\columnwidth]{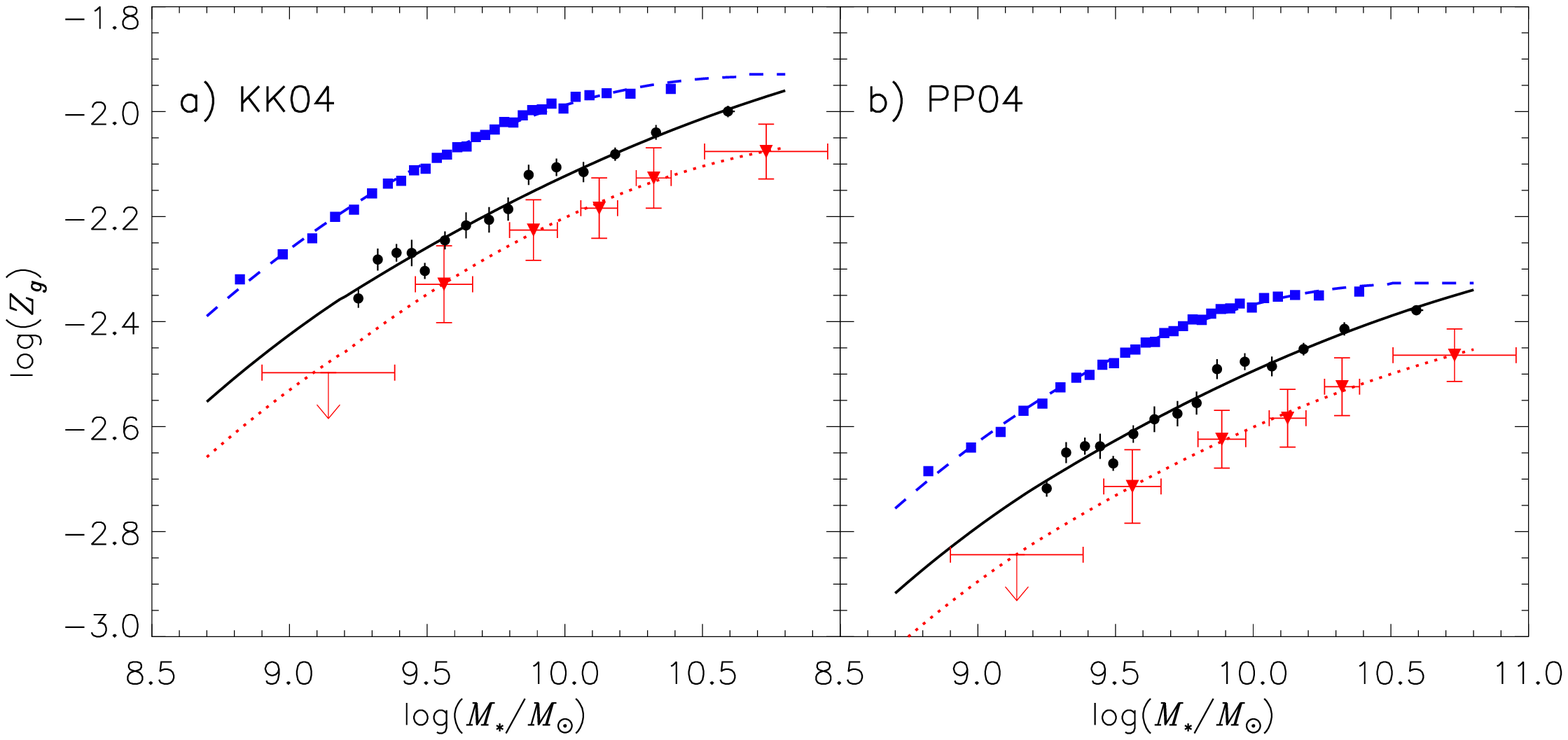}
\end{center}
\caption{The MZ relation determined using the calibration of a) \citet{Kobulnicky2004} and b) \citet{Pettini2004}. The metallicity is given as the mass abundance of oxygen relative to hydrogen. The blue, black and red curves are the MZ relations determined for the local sample from SDSS, the intermediate redshift sample from DEEP2 \citep{Zahid2011a} and the high redshift sample from \citet{Erb2006b}, respectively. The metallicities for the SDSS and DEEP2 data are originally determined using the \citet{Kobulnicky2004} diagnostic and the metallicity of the \citet{Erb2006b} sample is originally determined using the \citet{Pettini2004} diagnostic. When necessary, the metallicities have been converted using the coefficients given by \citet{Kewley2008}.}
\label{fig:zg}
\end{figure*}

The three data samples used in this study have been selected because they all have well determined MZ relations. The MZ relation for the SDSS and DEEP2 samples are determined by \citet{Zahid2011a} and for the E06 sample by \citet{Erb2006b}. We convert the metallicity, traditionally quoted as a number density given by 12 + log(O/H), to gas-phase mass abundance using the simple linear conversion given by Equation \ref{eq:mass_density}. 

In Figure \ref{fig:zg} we plot the MZ relation for our three data samples. The MZ relation for the SDSS and DEEP2 sample are determined using the calibration of \citet[Figure \ref{fig:zg}a]{Kobulnicky2004} and the for the E06 sample using the calibration of \citet[Figure \ref{fig:zg}b]{Pettini2004}. \citet{Kewley2008} have shown that while various calibrations give relatively accurate measurements of the metallicity, there are systematic differences. We apply the conversions constants they derive in order to convert between the two calibrations shown in Figure \ref{fig:zg}. The shape of the MZ relation determined from the two calibrations is very similar. The main difference being the 0.35 dex higher abundances of the \citet{Kobulnicky2004} calibration with respect to the calibration of \citet{Pettini2004}. In this study we adopt the average of the two calibrations for our determination of metallicities. In Section 8.3 we provide a more detailed discussion of the calibrations uncertainties.

We use the observed MZ relations to determine the gas-phase mass abundance as a function of stellar mass and redshift. The three samples in this study suggest that the MZ relation evolves linearly with time. \citet{Moustakas2011} investigate the MZ relation evolution out to $z=0.75$ and suggest that it evolves linearly with redshift. In this study we assume linear evolution with time but note that assuming linear evolution with redshift has only minor quantitive effects on our results and has no effect on our interpretation. We further assume a constant offset between the three relations at $\mathrm{log}(M_\ast/M_\odot) < 9.2$. The MZ relation in the local universe is shown to extend to low stellar masses \citep{Lee2006, Zahid2012} and we linearly extrapolate the three relations to lower stellar masses when necessary.

In summary, we determine the gas-phase mass abundance of oxygen at arbitrary redshifts and stellar masses by linearly interpolating between the three observed relations shown in Figure \ref{fig:zg}. We linearly interpolate both in \emph{time} and \emph{stellar mass} in order to determine $Z_g(M_\ast, z)$.

\section{A Census of Oxygen}
\label{sec:census}


There are three components to our oxygen census: total oxygen produced, amount of oxygen locked up in stars and the amount of oxygen in the gas-phase. To put this in context, we begin our census by noting the basic balance equation for oxygen. In a closed-box system with no inflows or outflows of gas, the oxygen balance is given by 
\begin{eqnarray}
dM_{g}^{o} & = & P_o \, \Psi - Z_{g} \,  \Psi + R\,Z_{g}\, \Psi \nonumber \\
& = & P_o \, \Psi - (1 - R) \,Z_{g}\, \Psi
\label{eq:chem_ev}
\end{eqnarray}
As before, $M_g^o$ is the mass of oxygen in the gas-phase, $Z_g$ is the gas-phase mass abundance of oxygen and is defined as $Z_g = M_g^o/M_g$, where $M_g$ is the gas mass. $\Psi$ is the SFR. $R$ and $P_o$ are the gas return fraction and mass of newly synthesized oxygen per unit stellar mass, respectively. The first term on the right-hand side of Equation \ref{eq:chem_ev} represents the total mass of oxygen produced, the second term represents the amount that goes into stars and the third term represents the amount that is subsequently returned through mass loss. The stellar yield, $y$, is related to the nucleosynthetic yield and gas fraction by $y = P_o/(1-R)$. We introduce this basic balance equation to place the census of oxygen in star-forming galaxies that follows within the context of simple chemical evolution models. For a more detailed discussion of these chemical evolution models we refer the reader to \citet{Edmunds1990}.

\subsection{Total Oxygen Production}

The total amount of oxygen produced by the stars within a galaxy is simply given by
\begin{equation}
M_T^o = P_o \, \frac{M_\ast}{1-R}.
\label{eq:ox_tot}
\end{equation}
The quantity $M_\ast/(1-R)$ represents the total amount of gas converted into stars. Several simplifying assumptions go into Equation \ref{eq:ox_tot}: 1) The IMF is constant, 2) gas is instantaneously returned and 3) the oxygen yield is independent of metallicity. In Section 8 we discuss systematics and uncertainties associated with each of these assumptions.

\subsection{Oxygen in the Gas-Phase}


\begin{figure}
\begin{center}
\includegraphics[width=\columnwidth]{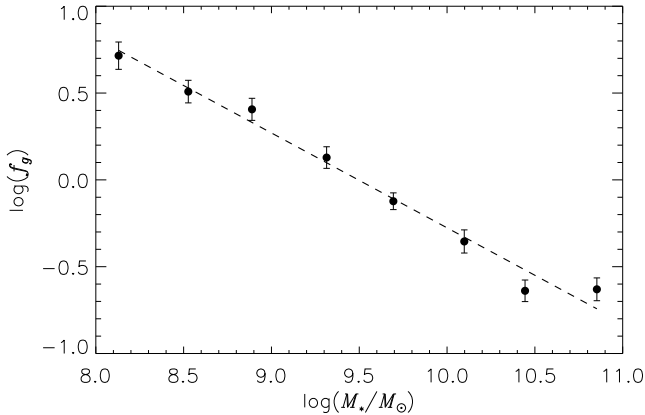}
\end{center}
\caption{The gas fraction as a function of stellar mass for star-forming galaxies taken from the compilation of \citet{Peeples2011}. The dashed line is a fit to the relation and is given by Equation \ref{eq:gf}.}
\label{fig:gf}
\end{figure}

The total mass of oxygen in the gas-phase in local star-forming galaxies is given by $M_g^o = Z_g \, M_g$, where $M_g^o$ is the mass of oxygen in the gas-phase. Unfortunately, direct gas mass measurements are available for only a few galaxies in our local sample. We therefore estimate the gas mass of local galaxies using the observed relation between gas fraction and stellar mass for star-forming galaxies. \citet{Peeples2011} compile gas mass estimates  (HI + H$_2$, with correction for helium) from several sources in the literature and give the binned cold gas fraction as a function of stellar mass (see Table 2 in their paper). The values are compiled from the data sets of \citet{McGaugh2005}, \citet{Leroy2008} and \citet{Garcia-Appadoo2009}. 

In Figure \ref{fig:gf} we plot the log of the gas fraction, $f_g$, as a function of stellar mass. The gas fraction is defined such that $f_g = M_g/M_\ast$\footnote{We have adopted the convention of \citet{Peeples2011} in defining $f_g$. This differs from the traditional definition of the gas fraction given by $f_g = M_g/(M_\ast + M_g)$.}. The stellar mass estimates of \citet{Peeples2011} are taken from MPA/JHU catalog and we make a 0.2 dex correction for consistency \citep{Zahid2011a}. The error bars are the $1\sigma$ uncertainties on the mean of log$(f_g)$. The dashed line is a fit to the relation given by
\begin{equation}
\mathrm{log}(f_g) = (-0.28 \pm 0.03) - (0.55 \pm 0.03)\cdot[\mathrm{log}(M_\ast/M_\odot) - 10].
\label{eq:gf}
\end{equation}
The errors in the parameters come from propagating the uncertainties in the mean of log$(f_g)$. When calculating gas mass, we add 0.1 dex to the fitted relation at all stellar masses to account for ionized gas \citep[see Table 1.1 in][]{Tielens2005}. Using this relation, the gas mass is given by $M_g = f_g M_\ast$. We note that the gas fractions given by Equation \ref{eq:gf} are greater than values inferred from inversion of the Schmidt-Kennicutt relation \citep[see Figure 2 in ][]{Peeples2011}. Using gas fractions determined from inverting the Schmidt-Kennicutt relation would therefore lower the estimate of the mass of oxygen in the gas-phase and increase the oxygen deficit presented in Section \ref{sec:odef}.

We can straightforwardly estimate the mass of oxygen in the gas-phase of local star-forming galaxies from observed quantities by combining the relation for $f_g$ given in Equation \ref{eq:gf} with the MZ relation for local star-forming galaxies shown in Figure \ref{fig:zg}. This is given by $M_g^{o} = Z_g f_g M_\ast$. 



\subsection{Oxygen Locked Up in Stars}

\begin{figure}
\begin{center}
\includegraphics[width=\columnwidth]{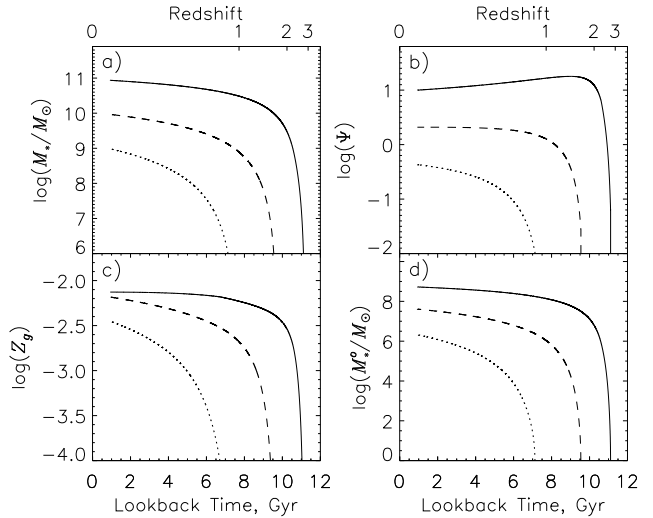}
\end{center}
\caption{The a) stellar mass, b) SFR ($M_\odot$ yr$^{-1}$), c) gas-phase oxygen abundance and d) the mass of oxygen locked up in stars determined from our models as a function of look back time and redshift. The solid, dashed and dotted curves are galaxies that have a stellar mass of $10^{11}, 10^{10}$ and $10^9 M_\odot$ in the local universe (z = 0.07). }
\label{fig:msz}
\end{figure}

The most difficult component of our oxygen census to estimate is the oxygen locked up in stars. The total amount of oxygen produced by stars and the amount of oxygen found in the gas-phase of galaxies can be inferred from the \emph{present-day} physical quantities of the stellar mass and gas-phase abundance. However, the oxygen locked up in stars is an accumulated property that is dependent on the star formation and chemical history. The amount of oxygen locked up in stars between some initial redshift $t_i$ and some later redshift $t$ is analytically given by
\begin{equation}
M^{o}_{\ast}(t>t_i) = (1 - R)\, \int^t_{t_i} Z_g(t^\prime) \, \Psi(t^\prime) \,dt^\prime.
\label{eq:mzsint}
\end{equation}
We have parameterized both the SFR (see Section \ref{sec:msr}) and gas-phase oxygen abundance (see Section \ref{sec:mzr}) as a function of stellar mass and time and are therefore able to empirically constrain the total amount of oxygen locked up in stars by simply carrying out a numerical integration of Equation \ref{eq:mzsint}. 

We simultaneously determine the star formation and stellar mass history of a galaxy using Equation \ref{eq:mint}. We set $M_{\ast,i} = 10^6 M_\odot$ and carry the integral from $t_i$ out to $t = 12.5$ Gyr \footnote{We set the upper limit of the integral to a time corresponding to a redshift of $z=0.07$ rather than $z = 0$ because this is the median redshift of our local sample from SDSS.}. Higher values of $t_i$ correspond to younger, lower stellar mass star-forming galaxies in the local universe. We show the stellar mass history of three galaxies ($t_i = 2.4, 4.0$ and 6.4 Gyr) in Figure \ref{fig:msz}a and star formation history in Figure \ref{fig:msz}b assuming $R=0.35$. The stellar mass history as a function of redshift allows us to determine the metallicity as a function of redshift for each galaxy using the procedure described in Section \ref{sec:mzr}. We plot the chemical history in Figure \ref{fig:msz}c. Entering the star formation history, $\Psi(t)$ (Figure \ref{fig:msz}b), and the chemical history, $Z_g(t)$ (Figure \ref{fig:msz}c), into Equation \ref{eq:mzsint}, we can determine the mass of oxygen locked up in stars as a function of redshift. This is shown in Figure \ref{fig:msz}d.

\begin{figure}
\begin{center}
\includegraphics[width=\columnwidth]{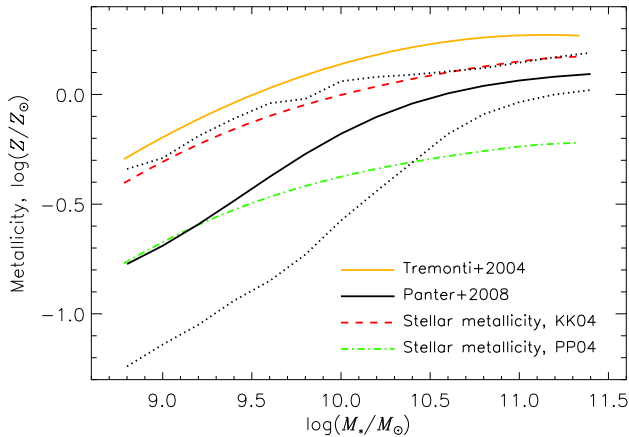}
\end{center}
\caption{The stellar mass$-$stellar metallicity relation for local star-forming galaxies. The solid black curve is the fitted relation determined by \citet{Panter2008} and the dotted curves are the 16 and 84\% contours of the distribution. The red dotted and green dot-dashed curves plot the stellar metal content of local star-forming galaxies calculated from Equation \ref{eq:stellar_metallicity} using the \citet{Kobulnicky2004} and \citet{Pettini2004} abundance calibrations, respectively. For reference \citep[c.f.][Figure 6]{Panter2008}, the solid yellow curve plots the MZ relation from \citet{Tremonti2004}.}
\label{fig:stellar_metallicity}
\end{figure}

\citet{Gallazzi2005} and \citet{Panter2008} examine the stellar mass$-$stellar metallicity relation determined from stellar population modeling of SDSS spectra. Both groups derive a consistent relation which implies that the imprint of the MZ relation from earlier epochs is observed in stellar populations of galaxies today. In Figure \ref{fig:stellar_metallicity} we compare the metallicity of stars from our model analysis to that observed by \citet{Panter2008} using $\sim300,000$ galaxies in SDSS. We estimate the mass-weighted stellar metallicity from our model of the star formation and chemical histories of galaxies (see Figure \ref{fig:msz}b and c, respectively). In particular we compute 
\begin{equation}
Z_\ast(M_\ast) = \frac{\int_{t_i}^{t} \Psi(t^\prime) Z(t^\prime) dt^\prime}{\int_{t_i}^{t} \Psi(t^\prime)dt^\prime} = \frac{\int_{t_i}^{t} \Psi(t^\prime) Z(t^\prime) dt^\prime}{M_\ast}, 
\label{eq:stellar_metallicity}
\end{equation}
where $Z_\ast$ is the mass-weighted stellar metallicity as a function of current stellar mass, $M_\ast$. We compare directly to the mass-weighted stellar metallicity relation observed and parameterized by \citet[see Equation 1 and Figure 6]{Panter2008}. In Figure \ref{fig:stellar_metallicity} the \citet{Panter2008} fitted relation is plotted by the solid black curve and the 16 and 84\% contours of the distribution are shown by the dotted black curve.

As can be seen from Equation \ref{eq:stellar_metallicity}, our estimate of the metal content of stars is dependent on the choice of gas-phase abundance diagnostic. In Figure \ref{fig:stellar_metallicity} we plot the mass-weighted stellar metallicity for both the \citet[red dashed curve]{Kobulnicky2004} and \citet[green dot-dashed curve]{Pettini2004} calibration. For reference the MZ relation from \citet{Tremonti2004} is plotted by the yellow curve and for consistency we adopt the solar metallicity value of 12+log(O/H)$_\odot = 8.87$ used by \citet{Panter2008}. Estimates from the two different abundance diagnostics converted into stellar metallicities using Equation \ref{eq:stellar_metallicity} envelope the relation from \citet{Panter2008}. In this study we have adopted a metallicity calibration that is an average of the \citet{Kobulnicky2004} and \citet{Pettini2004} calibration (see Section 7). Within the systematic uncertainties, our estimate of stellar metallicities using Equation \ref{eq:stellar_metallicity} and the observed stellar metallicities from \citet{Panter2008} are consistent.

\subsection{The Oxygen Deficit}
\label{sec:odef}
\begin{figure*}
\begin{center}
\includegraphics[width=2\columnwidth]{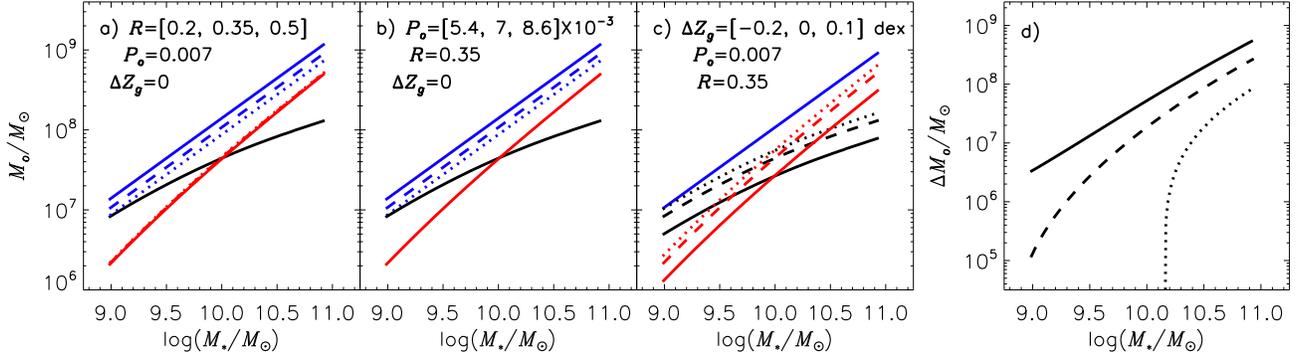}
\end{center}
\caption{The census model of oxygen determined by a) varying the return fraction, b) the oxygen yield and c) the zero point of the oxygen abundance calibration while keeping the other two parameters constant, respectively. a-c) The total oxygen produced (blue line), oxygen found in the gas (black) and stellar phase (red) are plotted as a function of stellar mass for local star-forming galaxies. In a-c) we display the three values adopted for the varying parameter along with the constant values adopted for the other two parameters. The solid, dotted and dashed lines indicate how the mass of oxygen varies for the three different census components when adopting the three different values for the varying parameter. d) The oxygen deficit which is defined as the total oxygen produced minus the oxygen found in the gas and stellar phase. The oxygen deficit is the same for the oxygen masses determined in a-c).}
\label{fig:do}
\end{figure*}

In Figure \ref{fig:do}a-c we plot the mass of oxygen for our three census components. We plot the total mass of oxygen produced (blue curve), the mass of oxygen in the gas-phase (black curve) and the mass of oxygen locked up in stars (red curve) as a function of stellar mass for star-forming galaxies in the local universe. The solid, dotted and dashed curves are used to illustrate the effect varying the parameters has on different components of the census. The three free parameters in our model are the return fraction, $R$, the oxygen yield, $P_o$, and a constant offset in metallicity (which we refer to as $\Delta Z_g$) since the zero point varies depending on the calibration. Figure \ref{fig:do}a-c demonstrates the dependencies of the various components of our census on the free parameters of the model. In Figure \ref{fig:do}a-c we vary $R$, $P_o$ and $\Delta Z_g$, respectively, while keeping the other two parameters fixed. The three values adopted for the varying parameter and the adopted values of the constant parameters are shown in a-c. In Figure \ref{fig:do}a we see that variations in $R$ mainly affects the total amount of oxygen produced with a small effect on the amount of oxygen locked up in stars. Figure \ref{fig:do}b shows that varying $P_o$ only affects the total amount of oxygen produced and Figure \ref{fig:do}c demonstrates that varying $\Delta Z_g$ affects both the stellar and gas-phase components of the census, but has no effect on the total amount of oxygen produced. 

For a closed-box model, the total mass of oxygen produced equals the mass of oxygen found in the gas-phase and stars. However if metals are lost from the system, the equality will not hold. Thus we have
\begin{equation}
\Delta M^o = M_T^o - M_g^o - M_\ast^o,
\end{equation}
where, as before, $M_T^o$ is the total mass of oxygen produced and $M_g^o$ and $M_\ast^o$ is the mass found in the gas-phase and stars, respectively. $\Delta M^o$, which we refer to as the oxygen deficit, represents the total amount of oxygen produced that is unaccounted for by the gas-phase and stellar components. In Figure \ref{fig:do}d we plot the oxygen deficit which is the same for the three cases shown in Figure \ref{fig:do}a-c. \emph{The variation in the parameters are degenerate such that the oxygen deficit depends only on the variation in the quantity $\Delta Z_g(1-R)/P_o$, effectively reducing our three free parameters to one free parameter}. We note that the stellar yield, $y$, is given by $y = P_o/(1-R)$. Therefore, our model is only sensitive to the changes in the zero point of the metallicity relative to the stellar yield. Higher (lower) values of $\Delta Z_g/y$ correspond to a smaller (larger) oxygen deficit. To first order, this is true for all models of chemical evolution relying on these quantities. 

If we assume that all oxygen in galaxies originates from the stars within the galaxy, the oxygen deficit cannot be negative because a negative value would imply that there is more oxygen in the gas and stars than was produced by the galaxy. A physical model of the oxygen deficit requires a combination of the free parameters ($\Delta Z_g (1 - R)/P_o$) that yield $\Delta M^o \geq 0$. The dotted line in Figure \ref{fig:do}d gives a negative oxygen deficit at $M_\ast \lesssim 10^{10} M_\odot$, implying this is an unphysical model. 

The functional form of the oxygen deficit with respect to stellar mass varies according to $\Delta Z_g(1-R)/P_o$. However, a generic feature of the oxygen deficit is that it monotonically increases with stellar mass and is substantial at high stellar masses regardless of our particular choice of $\Delta Z_g(1-R)/P_o$. In the case where the oxygen deficit is substantial at all stellar masses (solid curve in Figure \ref{fig:do}), we have $\Delta M^o \propto M_\ast^\alpha$ with a near unity value of for the exponent ($\alpha=1.14$). When the oxygen deficit is large at all stellar masses (solid curve in Figure \ref{fig:do}), the oxygen deficit does not strongly depend on the shape of the MZ relation or the shape of the relation between gas fraction and stellar mass. At more moderate values of $\Delta Z_g(1-R)/P_o$ examined in our models (dashed curve in Figure \ref{fig:do}), the relation between the oxygen deficit and stellar mass is more complicated. In Section 9.1 we argue that a model where the oxygen deficit is large at all stellar masses and is nearly proportional the stellar mass is most consistent with independent observations of oxygen in the halos of star-forming galaxies.

\begin{figure}
\begin{center}
\includegraphics[width=\columnwidth]{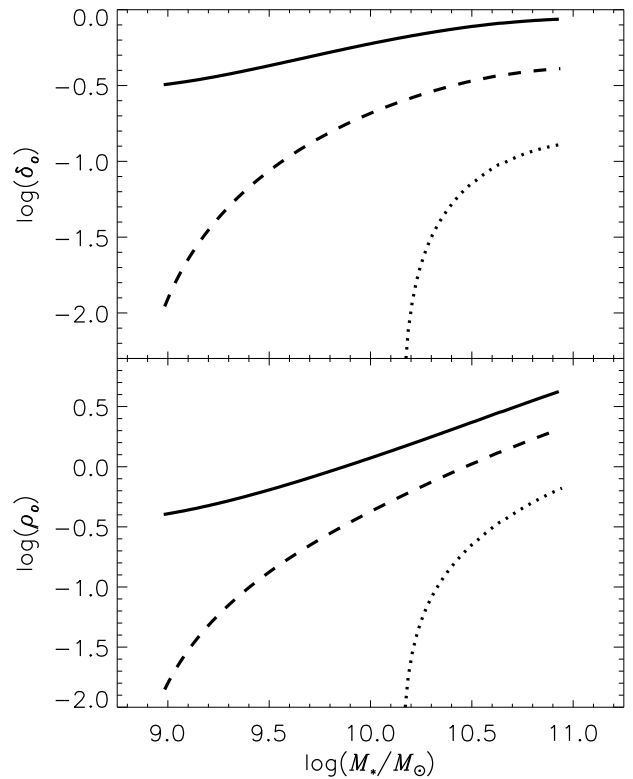}
\end{center}
\caption{The top panel shows the oxygen deficit relative to the mass of oxygen in the gas and stellar phase and the bottom panel shows the oxygen deficit relative to only the mass of oxygen in the gas-phase. The oxygen deficit is the same as in Figure \ref{fig:do}d and the solid, dotted and dashed lines indicate the different oxygen deficit derived by varying the free parameters.}
\label{fig:eta}
\end{figure}

In Figure \ref{fig:eta} we plot the fractional oxygen deficit. In the top panel we plot the logarithm of $\delta_o$, which we define as
\begin{equation}
\delta_o = \frac{\Delta M^o}{M_\ast^o + M_g^o}.
\end{equation}
$\delta_o$ is the oxygen deficit relative to the total oxygen content of the galaxy. In the bottom panel we plot the logarithm of $\rho_o$, which we define as 
\begin{equation}
\rho_o = \Delta M^o/M_g^o.
\end{equation}
$\delta_o$ is an important quantity reflecting key physical processes governing chemical evolution of galaxies but cannot be determined for individual galaxies as the fraction of oxygen locked up stars cannot be measured. $\rho_o$ does not account for oxygen locked up in stars, but can be determined from observable quantities and is therefore an important parameter for comparison with observations.

\section{Systematics and Uncertainties}
\label{sec:uncertainties}

We have derived the oxygen deficit by comparing the total oxygen produced with the amount found in the gas-phase and stars of galaxies. We have derived the oxygen deficit on the basis of observed relations that are determined from large samples of galaxies. Star formation rates and metallicities of individual galaxies are subject to stochastic variations resulting from physical processes that are not fully understood. The results of our study are valid in a statistical sense and do not necessarily apply to any individual galaxy.

Large stochastic variations in the star formation rates and metallicities of galaxies does pose a problem for our approach since we assume galaxies evolve along the observed MS and MZ relations. The tight relation between stellar mass and SFR and the small scatter suggest that the SFRs of star-forming galaxies in the local universe do not vary rapidly over their lifetime and that mergers play a minor role \citep[e.g.][]{Noeske2007a, Dutton2010, Elbaz2011}. Though the observed scatter in metallicities increases at the low stellar mass end of the local MZ relation \citep{Zahid2012}, the $1\sigma$ scatter is $\lesssim0.2$ dex \citep{Tremonti2004}. The main source of uncertainty then is likely not associated with our approach but rather with parameters of our model and other simplifying assumptions which we discuss below.

\subsection{The Return Fraction and Instantaneous Recycling Approximation}

The fraction of gas returned to the ISM is mass dependent and therefore sensitive to the particular choice of an IMF, with theoretical values ranging between $\sim20-50\%$. We have adopted this range in Figure \ref{fig:do}a. The physical mechanism regulating mass loss in stars varies by stellar mass. In low mass stars, stellar winds during the red-giant branch and asymptotic giant-branch are the dominant mechanism. In intermediate mass stars winds during the asymptotic giant branch phase dominate and for high mass stars, both stellar winds and supernovae contribute to stellar mass loss \citep[][and references therein]{Jungwiert2001}.

The return fraction is important in our models in that it sets the total amount of star formation required for a galaxy to have a particular current stellar mass. The total amount of gas converted into stars, $M_{gs}$, in a galaxy with a current stellar mass of $M_\ast$ is given by $M_{gs} = M_\ast/(1-R)$. A galaxy with $R = 0.5$ will, to first-order, form 40\% more stars than a galaxy with $R = 0.3$ to achieve the same stellar mass. Because the amount of oxygen produced is directly proportional to the total amount of star formation, there will be a commensurate increase in total oxygen production.

We have also adopted the instantaneous recycling approximation (IRA) introduced by \citet{Tinsley1980}. The IRA assumes that the gas is returned to the ISM immediately following star formation. It is shown that the IRA works well for high mass stars whose lifetimes are short compared to star formation time scales. Oxygen is a primary element, mainly produced in Type II supernovae. In our models, the lifetimes of massive stars responsible for oxygen enrichment and gas mixing timescales are significantly shorter than the timescales of star formation and therefore the IRA is reasonable in regards to oxygen production. However, the gas return from Type II supernovae accounts for 10-20\% of the gas returned to the ISM and a larger fraction is returned on longer timescales from intermediate and low mass stars \citep[see][]{Jungwiert2001}.

Examining various IMFs, \citet{Leitner2011} show that the majority of stellar mass loss occurs in the first 1-2 Gyrs (c.f. their Figure 1). \citet[among others]{Noeske2007a} argue that the MS relation implies that for most star-forming galaxies the SFR gradually declines. The gradual decline in SFRs is a generic feature of models assuming the continuity condition that galaxies build up stellar mass evolving along the observed MS relation \citep[e.g.][this work]{Peng2010, Leitner2011, Leitner2012}. The IRA can be reasonably applied in situations where the SFR is not changing rapidly with redshift because though the current generations of stars being formed will return majority of their gas in 1-2 Gyrs, the total gas return will be an integration over previous generations. If the SFR is not rapidly varying, the return fraction will reach a relatively constant value within a few Gyrs.

In the Appendix we apply a time dependent treatment for gas return in determining the oxygen deficit. We no longer assume the IRA so that oxygen is not instantaneously recycled nor is the return fraction a constant. We show that our results are not sensitive to time dependencies in the return rate and adopting a constant and instantaneous return of gas is valid.


\subsection{The Oxygen Yield}

The production of oxygen in our models is defined by the oxygen yield, $P_o$. The oxygen yield represents the total amount of newly synthesized oxygen per unit stellar mass of star formation. Following \citet{Henry2000}, we formally define the integrated oxygen yield as 
\begin{equation}
P_o = \int_{m_l}^{m_u} m p_o(m) \phi(m)dm.
\label{eq:int_yield}
\end{equation}
Here $m_l$ and $m_u$ are the lower and upper limits of the mass range of stars formed, respectively, $p_o(m)$ is the mass fraction of oxygen synthesized by a star of stellar mass $m$ and $\phi(m)$ is the normalized IMF. Uncertainties in $P_o$ arise from uncertainties in the stellar yields, the IMF and the mass range of integration. Here we discuss only the stellar oxygen yields. For a more complete review of uncertainties in stellar yields of various elements see \citet{Romano2010}.

The dependence of the oxygen yield on the IMF is clear from examination of Equation \ref{eq:int_yield}. The relative weight of stars in different mass ranges assuming different IMFs are given by \citet[][Table 1]{Romano2005}. The relative weight of $8 - 40 M_\odot$ stars is a factor 1.6 higher for the Chabrier IMF as compared to the Salpeter IMF, highlighting the uncertainties in the oxygen yields associated with the assumption of the IMF. These uncertainties are further exacerbated by the possibility of a varying IMF. \citet{Kroupa2003} argue that the integrated galactic initial mass function (IGIMF), the stellar mass distribution function for all stars within a galaxy, is an integration of the stellar IMF over the embedded cluster mass function, ECMF. There exists an empirical relation between the ECMF and the observed SFR, leading to a dependence of the IGIMF on the observed SFR. \citet{Peeples2011} show several models for the stellar yield, taking into consideration these variations.

Almost all oxygen in the universe is produced in massive stars ($m \gtrsim 8 M_\odot$). Using detailed models of Type II supernovae explosions, \citet{Woosley1995} provide the nucleosynthetic yields for stars in the mass range of $11 - 40 M_\odot$ and for a range of metallicities. \citet{Thielemann1996} present similar model calculations but in the mass range of $13 - 25 M_\odot$. \citet{Thomas1998} compare these two models finding good agreement between the oxygen yields except at high stellar masses where the yields calculated by \citet{Woosley1995} saturate. \citet{Henry2000} calculate the integrated oxygen yield using the models of \citet{Maeder1992}, \citet{Woosley1995} and the \citet{Thielemann1996} models updated to include contributions from $40$ and $70 M_\odot$ stars \citep{Nomoto1997}. Assuming a Salpeter IMF and a range of upper and lower masses \citet{Henry2000} find that the oxygen yields range between $0.004 - 0.016$ (see their Table 2). Using more sophisticated models, \citet{Kobayashi2006} calculate nucleosynthetic yields by assuming a range of metallicities, explosion energies (supernovae and hypernovae) and metallicity dependent mass loss. They calculate the integrated oxygen yield using a Salpeter IMF and a lower and upper mass limit of $0.07$ and $50 M_\odot$, respectively. They find that the oxygen yield is weakly dependent on metallicity and ranges between $0.006 - 0.008$. In Figure \ref{fig:do}b, we adopt a range of $0.0056 - 0.0091$ for the oxygen yield.


\subsection{The Abundance Calibration}

The uncertainty in the absolute nebular abundance calibration is a long-standing problem in observational astronomy. \citet{Kewley2008} show that the metallicity can vary by as much as 0.7 dex when using different abundance diagnostics for the same set of galaxies. This poses serious problems for understanding galactic chemical evolution. The uncertainties are largely due to method of calibration used in establishing the diagnostic. Empirical methods rely on calibrating strong line ratios using HII regions and galaxies with metallicities determined using the so-called direct method which uses temperature sensitive auroral lines. \citep[e.g.][]{Pettini2004}. Empirical methods have several limitations and theoretical calibrations of strong-line ratios using photoionization models have also been developed \citep[e.g.][]{Kewley2002, Kobulnicky2004, Tremonti2004}. Detailed discussions of strengths and weaknesses of the various methods and calibrations can be found in \citet{Kewley2008} and \citet{Moustakas2010}, we summarize the salient points here.

\citet{Peimbert1967} points out that temperature fluctuations within nebular regions may lead to overestimates of the temperature. Temperature fluctuations are thought to be more problematic in metal-rich HII regions, where efficient line-cooling may lead to temperature inhomogeneities and strong temperature gradients \citep{Garnett1992}. These overestimates of nebular temperatures lead to underestimates of the metallicity with the direct method \citep{Stasinska2002, Stasinska2005, Bresolin2006} and temperature inhomogeneities among the HII regions within a galaxy also pose problems for metallicity determinations based on global spectra \citep{Kobulnicky1999}. A second, and perhaps more pernicious problem is the distribution of electrons within an HII region may not follow a Boltzmann distribution but rather may be $\kappa$-distributed (a well known particle distribution in plasma physics). A $\kappa$-distribution of electrons may account for the discrepant temperatures and metallicities inferred from auroral lines \citep{Nicholls2012}.

Empirical calibrations require observations of extremely faint auroral lines. In particular, the [OIII]$\lambda4363$ line is on the order of 100 times weaker than the strong optical oxygen lines (e.g. [OII]$\lambda3727$, [OIII]$\lambda5007$). Furthermore, the line strength diminishes with increasing metallicity, due to efficient line cooling, and is only observed in low metallicity HII regions ($\lesssim0.5 Z_\odot$). The high S/N spectra required to observe the auroral lines taken together with temperature inhomogeneities within individual HII regions suggests that samples used to calibrate empirical methods may be biased and unreliable, particularly at higher metallicities.

Theoretical methods rely solely on photoionization models to calibrate strong-line ratios. These methods are not susceptible to observational limitations imposed by empirical calibrations and metallicities are well constrained and parameterization is well defined over the full range of observed line ratios. The absolute calibration of metallicities are model dependent and uncertainties are subject to the simplifying assumptions of the model. Indirect evidence suggests that these methods systematically overestimate the metallicity \citep[e.g.][]{Bresolin2009b, Kudritzki2012}. 


The metallicities in our samples are determined using either the theoretical diagnostic of \citet[SDSS and DEEP2]{Kobulnicky2004} or the empirical diagnostic of \citet[E06]{Pettini2004}. These two methods differ by a constant offset of $\sim0.3$ dex. For our metallicity diagnostic we therefore adopt the average of the metallicities determined using these two methods. 

\citet{Kewley2008} have shown that the relative estimates of the metallicity are robust for various diagnostics. However, the absolute calibration varies significantly and the shape of the MZ relation can change depending on the calibration adopted \citep[see Figure 2 of][]{Kewley2008}. In our models, the oxygen deficit is sensitive to the shape of the MZ relation only in the case when the oxygen deficit is small. As we show in the following section, our model showing a large oxygen deficit (the solid line in Figure \ref{fig:do}d) is most consistent with observations of oxygen in the halos of star-forming galaxies. Furthermore, the shapes of the MZ relations determined from the calibration of \citet{Kobulnicky2004} and \citet{Pettini2004} are consistent within the errors (see Figure \ref{fig:zg}) despite the fact that the two diagnostics are calibrated independently using a theoretical and empirical calibration, respectively.

\subsection{Depletion onto Dust Grains}

Some uncertainty in the absolute metallicity scale is associated with depletion of metals onto dust. Several authors have inferred $\sim0.1$ dex depletion of oxygen onto dust grains \citep{Jenkins2004, Cartledge2006, Peimbert2010}. However, in NGC 300 the agreement between the gradient inferred from A and B supergiants and that from emission lines suggests that there is very little, if any dust depletion in this galaxy \citep{Bresolin2009b}. It is likely that dust depletion ranges between $0\sim0.1$ dex and perhaps is dependent on physical properties of galaxies such as metallicity or stellar mass. 

The oxygen deficit we estimate is sensitive to the gas-phase abundance only at small oxygen deficits. If we assume a constant relative level of depletion (e.g. $\sim0.1$ dex), the effects of dust depletion on the oxygen deficit is more pronounced in lower mass galaxies which have a smaller oxygen deficit. However, low mass galaxies may also have lower levels of depletion as NGC300\footnote{The stellar mass of NGC300 is $\sim10^{9}M_\odot$ \citep{Kent1987}.} indicates. By comparing to independent observations of the mass of oxygen found in the halos of star-forming galaxies, we argue in Section 10.1 that a model with a large oxygen deficit is favored. In the case of a large oxygen deficit (solid curve in Figure \ref{fig:do}), assuming that oxygen is depleted onto dust by 0.1 dex changes the oxygen deficit by 0.7 dex at a stellar mass of $10^{9}M_\odot$ and by 0.15 dex at a stellar mass of $10^{11}M_\odot$. This level of dust depletion in our model with a large oxygen deficit does not alter the conclusions of this paper. Given the large uncertainties on how dust depleted oxygen is in HII region and systematic uncertainties of our metallicity calibration, we make no explicit correction for this effect. In Figure \ref{fig:do}c, we vary the adopted metallicity by [-0.2, +0.1] dex.


\subsection{Abundance Gradients}

It is well established that star-forming galaxies in the local universe have spatial abundance gradients such that the central regions of galaxies are more metal rich than their outskirts \citep[e.g.][]{Zaritsky1994a}. In our estimate for the amount of oxygen in the gas-phase (see Section 8.2) we adopt an oxygen abundance determined from nebular emission within the 3 arc second SDSS fiber aperture. We determine the oxygen abundance of local galaxies from data selected to have a covering fraction of $>30\%$ which is shown to be sufficient for reproducing the global abundance \citep{Kewley2005}. We reproduce the global luminosity weighted abundance whereas an unbiased method for estimating the gas-phase metal content of galaxies would be to use a global metallicity estimator that reflects the average total gas-phase abundance (i.e. a gas mass weighted abundance). However, given current observations it is not possible to assess what biases are introduced by adopting a luminosity weighted abundance for estimating gas-phase metal content. Surveys of abundance gradients in nearby galaxies currently under way along with a new generation of radio instruments capable of measuring the gas content of large samples of galaxies will likely resolve this issue.

We speculate that the luminosity weighted oxygen abundance is likely to overestimate the gas mass weighted abundance. In star-forming galaxies where luminosity is dominated by young stars the luminosity weighted abundance reflects the abundance of star-forming gas. The Schmidt-Kennicutt relation for star formation relates the star formation surface density to the gas surface density by $\Sigma_{SFR} \propto \Sigma_{gas}^{1.4}$ \citep{Kennicutt1998a}. Because of the non-linear relation between the two quantities, the luminosity weighted metallicity will not be consistent with the gas mass weighted metallicity but instead will be biased towards higher gas surface density regions which tend to lie in the central, metal-rich regions of galaxies. Therefore, the average metallicity of the gas is likely to be lower than the inferred luminosity weighted abundance used in this study for estimating the oxygen mass in the gas-phase. However, given that observations comparing the gas mass and luminosity weighted abundances are currently not feasible, we do not attempt to make a correction for this effect. 

\section{Discussion}
\label{sec:discussion}

The census of oxygen for star-forming galaxies has been determined from the observationally constrained star formation and chemical histories. In Section \ref{sec:census} we derive the oxygen deficit determined by comparing the expected total production of oxygen with what is currently found in the gas and stellar phases. We note that oxygen deficit represents the \emph{net} deficit of oxygen accounted for by the stars and gas and excludes oxygen that may be expelled from the ISM and subsequently recycled back into the galaxies as some models predict \citep[e.g][]{Dave2011b}. 

One advantage to determining mass loss in the manner presented in this study is that a detailed physical mechanism governing mass loss is not required. The feedback processes thought to be responsible for mass loss have proven to be notoriously difficult to model and disentangling the various contributions observationally is challenging given that the physical mechanisms are not clearly understood \citep[for review see][]{Veilleux2005}. Here we discuss some of the implications of our observational constraints on mass loss and future prospectives for our models and results.

\subsection{Outflows}

Galactic outflows are a key physical process governing galaxy evolution \citep{Veilleux2005}. In the nearby universe they are commonly observed in starburst and post-starburst galaxies \citep[e.g][]{Rupke2005, Martin2006, Tremonti2007, Rich2010, Tripp2011} and are ubiquitous in higher redshift star-forming galaxies \citep[e.g][]{Shapley2003, Weiner2009, Rubin2010, Steidel2010}. The outflowing gas and metals may remain gravitationally bound to  galaxies in which case the material should be found to reside in the hot halos or it may escape the galaxy potential well altogether.

The oxygen deficit may be partially resolved by oxygen present in the outer disks and hot halos of galaxies. Several studies have found flat abundance gradients in the outer disks of star-forming galaxies \citep{Bresolin2009a, Werk2010, Werk2011, Bresolin2012}. The amount of oxygen observed in the outer disks cannot be reconciled with the low levels of star formation and transport of metals from the inner disk is the most likely scenario explaining the flat abundance gradients though the physical mechanism for the transport of metals to outer disks is not clearly understood. Oxygen has also been observed in absorption in the halos of star-forming galaxies \citep{Chen2009, Prochaska2011}. In a survey of 42 galaxies conducted using the \emph{Cosmic Origins Spectrograph} onboard the Hubble Space Telescope, \citet{Tumlinson2011} report ubiquitous detection of OVI in the halos of star-forming galaxies. They conclude that the mass of heavy elements and gas in the circum-galactic medium (CGM) may far exceed whats found within the galaxies themselves. From the column density of OVI, they estimate that the CGM of $10^{9.5} - 10^{11.5}M_\odot$ stellar mass galaxies (out to 150 kpc) contains at least $\sim10^7 M_\odot$ of oxygen, where they have assumed an ionization correction factor of 0.2 to account for oxygen in all ionization states. 

\begin{figure}
\begin{center}
\includegraphics[width=\columnwidth]{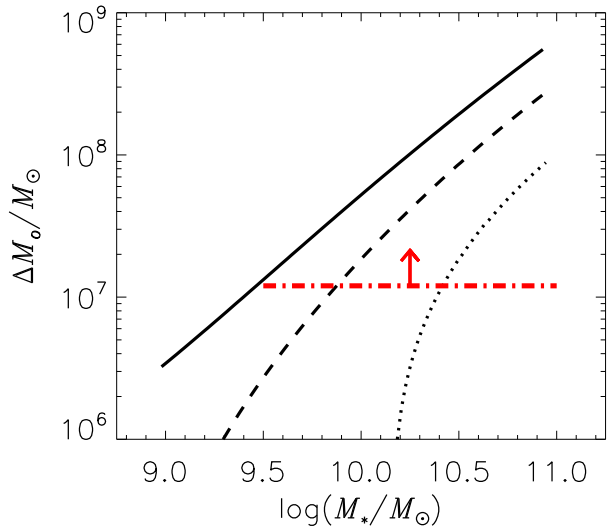}
\end{center}
\caption{The oxygen deficit with the lower limit of observed oxygen mass in the halos of star-forming galaxies from \citet{Tumlinson2011}. The oxygen deficit is the same as in Figure \ref{fig:do}d and the solid, dotted and dashed lines indicate the different oxygen deficit derived by varying the free parameters.}
\label{fig:do_oplot}
\end{figure}

In Figure \ref{fig:do_oplot} we plot the oxygen deficit along with the lower limit mass estimate of \citet{Tumlinson2011} shown by the red dot-dashed line. At a stellar mass of $10^{9.5} M_\odot$, our model showing the largest oxygen deficit (solid curve) is consistent with the lower limit estimates of \citet{Tumlinson2011}. These independent results rule out all except the highest oxygen deficit model (solid line in Figure \ref{fig:do_oplot}). \emph{A model with significant and ubiquitous outflows in star-forming galaxies is required for consistency with independent observations.} As discussed in Section 7.4, our models suggest that in the case of significant outflows the oxygen deficit will scale (nearly) linearly with stellar mass. A linear scaling is consistent with the result found by \citet{Kirby2011} from their analysis of metal mass loss in Milky Way dwarf galaxies. At $10^{11}M_\odot$, our model exceeds the lower limit estimate of \citet{Tumlinson2011} by $\sim2$ orders of magnitude.

Gas propelled to high velocities may escape from galaxies all together. Estimates of the escape fraction of outflowing material have been difficult to determine accurately mainly due to lack of constraints on halo drag \citep{Veilleux2005}. It may be that detailed estimates of the oxygen mass in the CGM of galaxies will account for the $\sim2$ orders of magnitude greater oxygen deficit at $M_\ast \sim 10^{11} M_\odot$ implied by our models as compared to lower limit estimates of \citet{Tumlinson2011}. However, if the oxygen deficit is not resolved by oxygen found in the halos of galaxies, a natural reservoir for the remainder of the oxygen deficit would be the intergalactic medium (IGM). This scenario is consistent with theoretical models where large scale galactic outflows transport metals from the galaxy to the IGM \citep[e.g.][]{Springel2003a, Oppenheimer2006, Kobayashi2007, Oppenheimer2011}. However, the kinematics of oxygen in the CGM of local star-forming galaxies suggest that most of the oxygen is gravitationally bound \citep{Tumlinson2011}.

A common method of estimating the escape fraction is to compare the outflow velocity to the escape velocity of the gravitational potential well of the galaxy. Early theoretical work suggested that mass loss occurs only in galaxies with shallow potential wells \citep[e.g.][]{Larson1974, Dekel1986, MacLow1999, Ferrara2000}. However, recent observational and theoretical work has suggested that mass loss driven by AGN or star formation occurs in significantly higher mass galaxies as well \citep{Strickland2004, Murray2005, Weiner2009, Steidel2010, Fischer2010, Feruglio2010, Sturm2011, Aalto2012}. 

It is not clear whether low-mass or high-mass galaxies are responsible for the IGM enrichment. Though low mass galaxies may lose a larger fraction of their metals, they may not be a significant source of IGM enrichment due to their low rates of metal production \citep{Martin2002, Kirby2011}. On the other hand, due to the deep potential wells of large galaxies, metals may not be efficiently ejected into the IGM. If the latter is true, our model would predict that the CGM and/or outer disks of massive star-forming galaxies ($10^{11}M_\odot$) should contain substantially greater quantities of oxygen than their lower mass counterparts ($10^9M_\odot$). Surveys of the CGM and studies of the outer disks of star-forming galaxies currently underway will provide important clues to resolving the oxygen deficit implied by our models.




\subsection{Baryonic Mass Loss}

One possible interpretation of the oxygen deficit is that it has been driven out of the galaxy ISM by galactic outflows. If oxygen is expelled via outflows, we expect an even larger mass of gas expelled since outflowing gas will not be pure oxygen. We can estimate the \emph{total} baryonic mass loss in two ways. First, we adopt an enriched wind model assuming that the metallicity of outflowing gas is equal to the Type II SNe oxygen yield (see Section 8.2). We adopt a value for $P_o = 0.007$ which is $\sim1.6Z_\odot$. This value is consistent with estimates of \citet{Martin2002} for the metallicity of a starburst driven wind in NGC 1569. Assuming outflows are enriched, we can estimate the total baryonic mass loss by the dividing the oxygen deficit by the oxygen yield. This is given by
\begin{equation}
\Delta M_b = \frac{M_T^o - M_g^o - M_\ast^o}{P_o} = \frac{\Delta M_o}{P_o}.
\label{eq:db}
\end{equation}
$\Delta M_b$ is the total amount of baryonic mass loss and $\Delta M_o$ is the oxygen deficit.

\begin{figure}
\begin{center}
\includegraphics[width=\columnwidth]{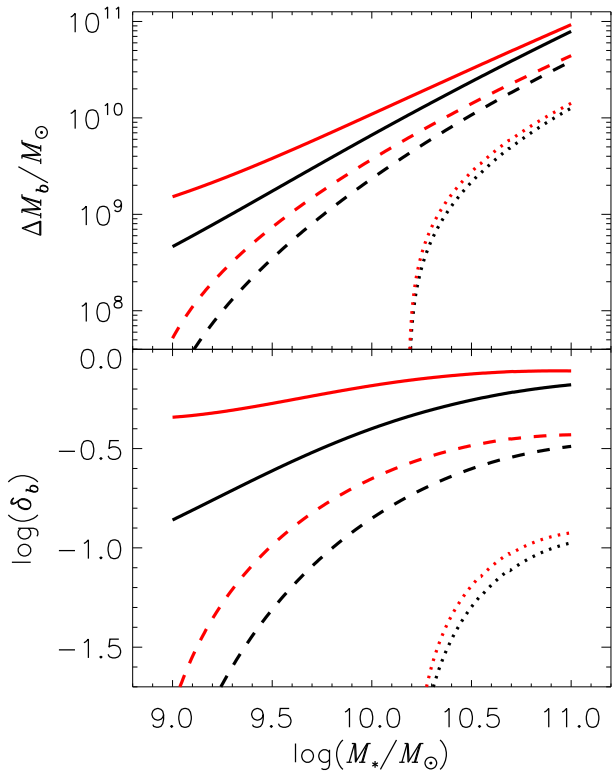}
\end{center}
\caption{The top panel is the inferred baryonic mass loss and the bottom panel is the baryonic mass loss relative to the stellar and gas mass. The oxygen deficit is the same as in Figure \ref{fig:do}d and the solid, dotted and dashed lines indicate the different oxygen deficit derived by varying the free parameters. The black and red curves are determined by adopting an enriched outflow (Equation \ref{eq:db}) and uniform wind model (Equation \ref{eq:dbupper}), respectively.}
\label{fig:db}
\end{figure}

We also derive an upper limit estimate of the total baryonic mass loss by assuming a uniform wind model. In the uniform wind model, the metallicity of the outflowing gas is the same as the ISM. In general, the metallicity of the outflowing gas is lower than the yield and therefore more gas is required to outflow in this model in order to yield the oxygen deficit derived in Section 7.4. In the uniform wind model estimate $P_o$ in Equation \ref{eq:db} is replaced by a SFR weighted metallicity, $Z_\psi$. The SFR weighted metallicity and the total baryon loss are given by
\begin{equation}
Z_\psi = \frac{\int_{t_i}^{t} \Psi Z_g dt}{\int_{t_i}^{t} \Psi dt}
\end{equation}
and
\begin{equation}
\Delta M_{b\psi} = \frac{\Delta M_o}{Z_\psi},
\label{eq:dbupper}
\end{equation}
and $\Delta M_{b\psi} > \Delta M_b$.

In the top panel of Figure \ref{fig:db} we plot the total baryonic mass loss as a function of stellar mass for galaxies in the local universe. The black and red curves are estimates adopting an enriched outflow (Equation \ref{eq:db}) and uniform wind model (Equation \ref{eq:dbupper}), respectively. In the bottom panel of Figure \ref{fig:db} we plot the logarithm of $\delta_b$ which we define as
\begin{equation}
\delta_b = \frac{\Delta M_b}{M_g + M_\ast}.
\end{equation}
$\delta_b$ is the ratio of the total baryonic mass loss to the baryonic mass of the galaxy, where the baryonic mass of the galaxy (gas + stellar). $\delta_b$ represents the ratio of the mass of gas to be cycled in and out of the galaxy compared to the current baryonic mass of the galaxy.

In the currently accepted cosmological model ($\Lambda$CDM), the universal baryon fraction is precisely determined from the cosmic microwave background, the observed baryon acoustic oscillations and the Hubble constant. The universal baryon and dark matter density revealed by these observations is given by $\Omega_b = 0.0456 \pm0.0016$ and $\Omega_c = 0.227 \pm 0.014$, respectively \citep{Komatsu2011}. Studies attempting to account for the baryons find that only a fraction of the expected baryons are observed in the low-redshift universe \citep{Fukugita1998, Fukugita2004, Nicastro2005, Sommer-Larsen2006, Shull2011}.

Only about a tenth of the baryons are found in the stars and gas of galaxies \citep{Bell2003a}. While the Ly$\alpha$ forest at low redshifts can account for another $\sim30\%$ \citep{Penton2004, Sembach2004}, the majority of baryons are still missing. Some cosmological simulations favor the warm-hot intergalactic medium (WHIM) as the repository of the missing baryons \citep[e.g][]{Cen1999, Dave2001, Cen2006, Oppenheimer2011}. However, observations of the hot gas at $10^5 - 10^7$K comprising the WHIM remain tentative and no compelling evidence for the detection of this phase yet exists \citep{Bregman2007}. The distribution of the WHIM material is unknown and hot halos of massive galaxies are considered as a possible reservoir for substantial fraction of the missing baryons \citep{Cen2006, Tang2009, Kim2009}, though this remains controversial \citep{Anderson2010}.

An open question is what fraction of the missing baryons were ejected from galaxies through feedback processes and what fraction never accreted in the first place. We can compare our estimates of the total baryonic mass that can be associated with galaxies (baryonic mass plus the baryonic mass loss) with the expected baryon content of galaxies and their halos inferred from cosmological estimates. The stellar-to-halo mass (SHM) relation parameterizes the relationship between stellar mass of galaxies and the dark matter halos in which they reside. \citet{Moster2010} develop a statistical approach whereby halos and subhalos are populated in an $N$-body simulations with the requirement that the observed stellar mass function be reproduced. They parameterize the SHM as 
\begin{equation}
\frac{M_\ast}{M_h} = 2 \left(\frac{M_\ast}{M_h}\right)_0 \left[ \frac{M_h}{M_1}^{-\beta} + \frac{M_h}{M_1}^{-\gamma} \right]^{-1}.
\end{equation}
Here $M_h$ is the halo mass and $\left(\frac{M_\ast}{M_h}\right)_0$, $M_1$, $\beta$ and $\gamma$ are free parameters. The relation evolves with redshift and the parameters are given by
\begin{eqnarray*}
\mathrm{log} M_1 (z) & =& 1.07 \cdot(1+z)^{0.019} \nonumber \\
 \left(\frac{M_\ast}{M_h}\right)_0 (z) &=& 0.0282 \cdot(1+z)^{-0.72} \nonumber \\
 \gamma (z) &=& 0.556 \cdot (1+z)^{-0.26} \nonumber \\
 \end{eqnarray*}
and
\begin{equation}
\beta (z) = 1.06 + 0.17 z. \nonumber
\end{equation}
We can estimate the expected baryon content of galaxies from the universal ratio of baryonic to dark matter given by $f_{bc} = \Omega_b/\Omega_c =  0.201 \pm 0.014$.

\begin{figure}
\begin{center}
\includegraphics[width=\columnwidth]{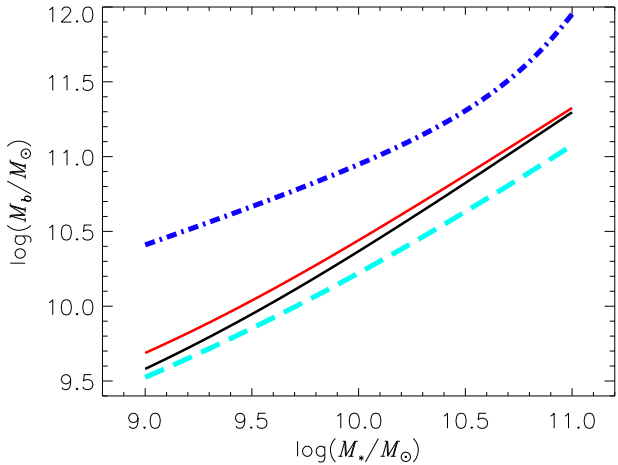}
\end{center}
\caption{The dashed cyan curve is the total baryonic (gas + stellar) mass of galaxies in the local universe. The dot-dashed blue line is the inferred baryon content from cosmological fraction and is given by $f_{bc} \, M_h$. The solid black and red lines are the total baryonic mass plus the baryon mass loss for an enriched wind model (Equation \ref{eq:db}) and uniform wind model (Equation \ref{eq:dbupper}), respectively. In our estimate we have adopted the oxygen deficit given by the solid curve in Figure \ref{fig:do}d.}
\label{fig:mb}
\end{figure}

Figure \ref{fig:mb} demonstrates the missing baryon problem for galaxies. The dashed cyan line is the current total baryonic (gas + stellar) mass content of galaxies plotted as a function of stellar mass. The dot-dashed blue line is the expected total baryonic mass content of galaxies assuming the universal baryonic to dark matter ratio and the SHM relation of \citet{Moster2010} and is given by $M_b = f_{bc} \, M_h$. The solid black and red curves are the sum of the baryonic mass and the baryonic mass loss using the two estimates given by Equations \ref{eq:db} and \ref{eq:dbupper}. The solid black and red curves can be interpreted as estimates of the baryon content associated with galaxies and are the amount of baryons currently found in local star-forming galaxies plus what was once in the galaxies but has since been cycled out through outflows. 

The baryon content associated with galaxies is still substantially lower than the inferred content from cosmology. The straightforward interpretation of Figure \ref{fig:mb} is that the missing baryons were never accreted onto local star-forming galaxies and unless a large reservoir of gas is found in the halos of star-forming galaxies it is likely that the missing baryons reside in the IGM. Using observationally motivated constraints for the mass and radii of hot halos, \citet{Anderson2010} come to a similar conclusion.

\begin{figure}
\begin{center}
\includegraphics[width=\columnwidth]{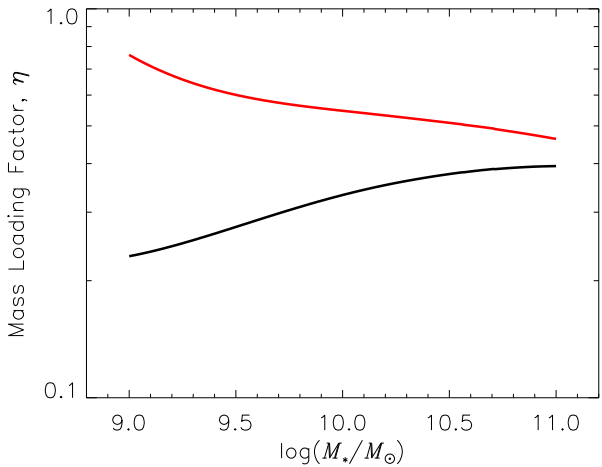}
\end{center}
\caption{The effective mass loading parameter given in Equation \ref{eq:ml} plotted against stellar mass. The red and black curves give the effective mass loading factor for our upper and lower limit estimates of the total baryon mass loss, respectively.}
\label{fig:ml}
\end{figure}

From our estimate of the baryon mass loss we can derive an effective mass loading factor, $\eta$, which is given by
\begin{equation}
\eta = \frac{\Delta M_b}{M_\ast/(1-R)}.
\label{eq:ml}
\end{equation}
The quantity $\Delta M_b$ is the total amount of baryonic mass loss and $M_\ast/(1-R)$ represents the \emph{total} amount of star formation. In the literature the instantaneous mass loading factor is defined as $\eta_i = \dot{M}_w/\dot{M_\ast}$. The instantaneous mass loading factor is the mass loss rate divided by the SFR. The effective mass loading as we have defined it in Equation \ref{eq:ml} is the instantaneous mass loading factor, $\eta_i$, averaged over the star formation history. In Figure \ref{fig:ml} we plot the effective mass loading factor as a function of stellar mass. Here, we have adopted $R=0.5$. Assuming a smaller value for $R$ would lower the estimate. Even adopting our upper limit estimates of baryon mass loss, the effective mass loading for star-forming galaxies in the local universe is $<1$.

\subsection{Future Prospects}

A complete and self-consistent theory of galaxy evolution will require a detailed account of galaxy growth and chemical evolution along with physical mechanisms governing these processes. In this contribution we present empirical models attempting to self-consistently integrate chemical evolution and galaxy growth. Our self-consistent approach is complementary to cosmological simulations and semi-analytical models and the self-consistent census approach presented here should be compared with those models. Both theoretical and observational advances are crucial to constraining the model results developed in this study. The method of analysis used in this study provides useful tests for consistency of a diverse set of observations and theories related to chemical properties of galaxies. Here we address future prospects for improvement.


One of the greatest outstanding astrophysical problems is the large uncertainty in the absolute calibration of the nebular abundances scale. High S/N observations of large sample of HII regions covering a broad range of physical parameters will be extremely important in statistically establishing and testing diagnostics. Such observations will be crucial in developing empirical calibrations relying on recombination lines which are thought to be less susceptible to effects of temperature variations associated with auroral lines \citep{Esteban2002, Peimbert2005, Bresolin2007}. A complementary approach will be to use the recently developed wide-field integral field spectrographs to observe nearby HII regions in order to understand the discrepancy between empirically and theoretically calibrated strong-line methods. A well calibrated diagnostic applied to large data sets investigating the MZ relation at higher redshifts and to lower stellar masses will provide important constraints for our understanding of chemical evolution and our census of oxygen in the stellar and gas-phase.

Theoretical models incorporating stellar rotation and mass loss into computations of nucleosynthesis for stars at all masses are yet to be developed \citep{Romano2010}. Current models of stellar nucleosynthesis are not yet able to explain the full diversity of chemical abundance patterns observed. While the oxygen yields are much better constrained by models owing to its primary origin, there is still a factor of $\sim2$ discrepancy. A single, self-consistent model of nucleosynthesis able to reproduce the abundance ratios observed in the Milky Way and other nearby galaxies will likely alleviate some of the discrepancy, thus constraining the total oxygen production in star-forming galaxies.

Part of the uncertainty in chemical evolution models rests on the assumption of a particular IMF \citep{Romano2005}. Adopting a constant IMF is problematic for galaxies, where even a constant stellar IMF, could lead to variations in the integrated galactic IMF \citep{Kroupa2003} and studies of the IMF from integrated measurements of star-forming galaxies indirectly indicate variations \citep{Hoversten2008, Lee2009, Meurer2009}. A varying IMF would have important implications for chemical evolution models \citep[e.g.][]{Romano2005, Koppen2007, Calura2010}. Despite the mounting evidence, no direct evidence of IMF variations in star-forming galaxies is currently available leading to a lack of consensus on the constancy of the IMF. Distinguishing whether a universal IMF or IGIMF formulation provides a better description of large scale star formation will be important for chemical evolution studies of galaxies. Large statistical studies of the integrated properties of galaxies observed over cosmic time, in particular observations of the FUV and FIR properties which are now becoming available, will be useful in establishing any systematic variations of the IMF. In our study, tighter constraints on the IMF will alleviate much of the uncertainty in the total amount of oxygen produced and the return fraction.

Total gas masses are required to measure the absolute content of metals within galaxies from observed relative abundances. In the local universe, large surveys of atomic and molecular gas have been conducted \citep{Helfer2003, Walter2008, Leroy2009, Saintonge2011}. Molecular gas at higher redshifts has also been detected in star-forming galaxies \citep{Daddi2008, Daddi2010, Tacconi2010}. The next generation of radio and sub-mm observatories such as the Atacama Large Millimeter/submillimeter Array and the Square Kilometer Array will revolutionize the study of cold gas in the universe allowing us to probe larger samples to far greater depths. Currently, we are only able to estimate the total oxygen in the gas-phase of local star-forming galaxies. Measurements of cold gas in nearby and distant star-forming galaxies taken together with a well calibrated gas-phase metallicity diagnostic will allow us to track the total mass, not just relative abundance, of oxygen in the gas-phase of star-forming galaxies over cosmic time.

The structure, content and chemical composition of outflowing gas is highly uncertain. Understanding the physical properties of galactic winds is crucial for estimating the total mass of gas outflowing from galaxies. Observations reveal that galaxy scale outflows have a complicated multiphase structure \citep[see][]{Veilleux2005} and observations suggest that structure and composition vary in each of the phases \citep[e.g.][]{Tripp2011}. Multi-wavelength observations are required. The cold gas component can be observed in absorption lines of neutral and low ionization state metals and the current generation of integral field spectrographs on 8-10m class telescopes and wide-field integral field spectrographs on 4m class telescopes will provide census of galactic scale outflows in local galaxies and presence of galactic winds in the distant universe. Hotter phases can be observed using space-based UV spectrometers such as the \emph{Cosmic Origins Spectrograph} (COS) onboard Hubble. A crucial but inaccessible phase is the so-called wind fluid which drives the stellar winds. A hard x-ray telescope with high spatial resolution and sensitivity is required to observe this phase. For the foreseeable future, astronomers will likely have to rely on detailed theoretical models and indirect observations to understand the wind fluid.

COS will also provide important insight into the circum-galactic and warm-hot ionized mediums, both thought to be important repositories of baryons. Studies of these regions will begin to reveal the baryonic content of the hot haloes and intergalactic medium surrounding galaxies. Observations of these regions along with a census of baryons associated with galaxies will be crucial to potentially identifying the large fraction of missing baryons in the local universe and resolving this long-standing problem \citep{Bregman2007}. A benchmark for these studies is to resolve the oxygen deficit in star-forming galaxies presented here.

In this study we have applied the best theoretical and observational constraints available in undertaking a census of oxygen in star-forming galaxies. We are unable to draw any strong quantitative conclusions from the models developed owing to the large uncertainties associated with both our adopted model parameters and observational inputs. Nonetheless, this study represents one of the first attempts to self-consistently account for the stellar mass growth and chemical evolution of galaxies. As theoretical and observational advances allow for ever greater constraints, we hope the approach taken in this study will prove to be an important ingredient in testing and developing a fully self-consistent theory of galaxy evolution.

\section{Summary}
\label{sec:summary}

In this study, we have consistently incorporated chemical evolution in the framework of stellar mass growth in order to conduct a census of oxygen in local star-forming galaxies. We are able to estimate the total oxygen production from total amount of star formation inferred from the current stellar mass. The mass of oxygen in the gas-phase is constrained by the observed MZ relation in the local universe and the relation between gas fraction and stellar mass. The most difficult to constrain observationally is the amount of oxygen locked up stars. Our empirical models self-consistently incorporate stellar mass growth and chemical evolution, thus allowing us to track the metallicity of the gas from which stars are formed over cosmological timescales and giving us empirical constraints on the oxygen mass locked up in stars. The main results of this study are given below.

\begin{enumerate}


\item{We conduct a census of oxygen and show that the amount of oxygen in the stellar and gas phase of galaxies does not fully account for the total amount of oxygen produced. We conclude that the most straight-forward interpretation of the oxygen deficit is that oxygen has been expelled from galaxies by outflows. Our results establish the need for ejective feedback in normal star-forming galaxies.}

\item{We compare our oxygen deficit with the observed lower limit of oxygen found in the CGM of star-forming galaxies and conclude that oxygen mass loss is a ubiquitous process in star-forming galaxies. Furthermore, the oxygen deficit in our preferred model scales with stellar mass and we predict that either more massive galaxies should be found to contain a greater mass of oxygen in their halos or that oxygen escapes the galaxy potential well altogether and therefore massive galaxies contribute for IGM enrichement.}

\item{We estimate the total amount of mass lost from the ISM of star-forming galaxies and find that it is a small fraction of the total baryon content expected from the cosmological baryon density. We conclude that only a small fraction of the total baryons in the universe ever cycled through star-forming galaxies.}

\end{enumerate}

Our empirical model provides an important test of self-consistency for many physical processes governing galaxy evolution. Future theoretical and observational advances will provide ever increasing constraints on the census of oxygen in star-forming galaxies and our models provide important benchmarks with which to compare theory and observation.

\acknowledgements

\emph{Acknowledgments}: We would like to thank the anonymous referee for useful comments that have improved the paper. HJZ and LJK gratefully acknowledge support by NSF EARLY CAREER AWARD AST07-48559. HJZ thanks members of the Extragalactic Discussion Group, conversations with whom lead to the ideas presented in this paper. We thank the DEEP2 team for making their data publicly available. The analysis pipeline used to reduce the DEIMOS data was developed at UC Berkeley with support from NSF grant AST-0071048. Funding for the SDSS and SDSS-II has been provided by the Alfred P. Sloan Foundation, the Participating Institutions, the National Science Foundation, the U.S. Department of Energy, the National Aeronautics and Space Administration, the Japanese Monbukagakusho, the Max Planck Society, and the Higher Education Funding Council for England. The SDSS Web Site is http://www.sdss.org/.

\section*{Appendix}
\setcounter{secnumdepth}{-1}
\subsubsection{A1. Time Dependent Mass Return}

\begin{figure}[b!]
\begin{center}
\includegraphics[width=\columnwidth]{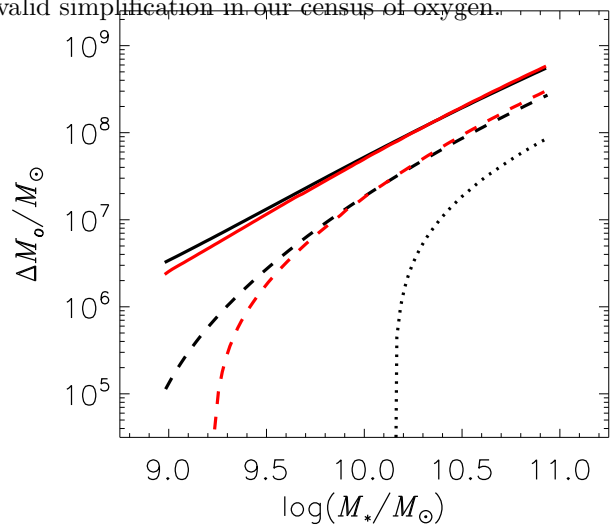}
\end{center}
\caption{The black curves are the same as in Figure \ref{fig:do}d. The red curves are the oxygen deficit determined by applying a time dependent return rate using the Chabrier (solid curve) and Chabrier steep (dashed curve) IMFs.}
\label{fig:return}
\end{figure}

Here we test the validity of our assumption of a constant return rate with instantaneous recycling. \citet{Jungwiert2001} develop an algorithm to account for continuous, time dependent mass loss. They give the gas return rate as
\begin{equation}
\dot{M}_R(t) = \int_0^t \Psi(t^\prime) \dot{f}_{mr}(t - t^\prime) dt^\prime. 
\label{eq:grr}
\end{equation}
Here $\dot{M}_R(t)$ is the gas return rate as a function of time and $\dot{f}_{mr}(t - t^\prime)$ is the mass loss rate at time $t$ for a single burst stellar population. The mass loss then is given as a convolution of the SFR with the mass loss rate (Equation \ref{eq:grr}). The gas return for various IMFs is well parameterized by the equation
\begin{equation}
\dot{f}_{mr}(t) = C_0 \, \mathrm{ln} \left(\frac{t}{\lambda} + 1 \right).
\label{eq:return}
\end{equation}
Here both $C_0$ and $\lambda$ are constants depending on the particular choice of IMF \citep[values are given in Table 1 of][]{Leitner2011}. We determine the gas return rate using a Chabrier and a Chabrier steep IMF. These two IMFs have been chosen because they are representative of the upper and lower mass loss rates of most IMFs \citep[see Figure 1 of][]{Leitner2011}.

We employ the similar algorithm for determining a self-consistent gas return rate and stellar mass history as outlined in \citet{Leitner2011}. We determine a star formation history assuming Equation \ref{eq:mint} with the return fraction $R = 0.35$. Using this star formation history we then determine the gas return rate from Equation \ref{eq:return}. We then replace the constant return fraction with the time dependent return rate in Equation \ref{eq:mdot} such that we have
\begin{equation}
\frac{dM_\ast}{dz} = \Psi - \dot{M}_R.
\end{equation}
We integrate this equation to determine a new star formation history which we then use to redetermine a new gas return rate via Equation \ref{eq:grr}. We iterate until $\Delta M_\ast < 0.01$, where $\Delta M_\ast$ is the change in the final stellar mass between iterations. This procedure typically takes $\lesssim10$ iterations. In this procedure, the gas returned has the metallicity of the natal gas, thus relaxing the assumption of instantaneous recycling.

Following the procedure outlined in Section \ref{sec:census}, we determine the oxygen deficit using the time dependent return rate formulation. The oxygen deficit is shown in Figure \ref{fig:return} where we have used $P_o = 0.008$ and $\Delta Z_g = 0$. The black curves are the same as Figure \ref{fig:do}d and are determined using a constant, instantaneous return rate. The solid and dashed red curves are the oxygen deficit determined from the Chabrier and Chabrier steep IMFs, respectively. This figure convincingly demonstrates that a constant return rate with instantaneous recycling is a valid simplification in our census of oxygen.

\bibliographystyle{apj}

\bibliography{manuscript}

 \end{document}